\documentclass[twocolumn]{aastex63}
\usepackage{booktabs}
\usepackage{IEEEtrantools}
\usepackage{graphicx}
\usepackage{enumitem}

\usepackage{amsmath}
\usepackage{comment}
\usepackage{tabularx}
\usepackage{multirow}
\definecolor{yscol}{rgb}{0.8, 0.6, 1}

\usepackage{natbib}
\usepackage{verbatim}
\usepackage{mathptmx}
\usepackage{url}
\usepackage{numprint}

\newcommand{\msun}{\,{\rm M}_\odot}
\newcommand{\pc}{\,{\rm pc}}
\newcommand{\kpc}{\,{\rm kpc}}
\newcommand{\km}{\,{\rm km}}
\newcommand{\s}{\,{\rm s}}
\newcommand{\g}{\,{\rm g}}

\newcommand{\yr}{\,{\rm yr}}

\definecolor{yscol}{rgb}{0.8, 0.6, 1}

\definecolor{yscol}{rgb}{0.8, 0.6, 1}

\received{November XX, 2022}
\revised{November XX, 2022}
\accepted{November XX, 2022}
\shorttitle{Growth of A Massive Black Hole Via Tidal Disruption Accretion}
\shortauthors{Seungjae Lee, Ji-hoon Kim, and Boon Kiat Oh}
\graphicspath{{./}{figures/}}
\begin{document}

\title{Growth of A Massive Black Hole In A Dense Star Cluster Via Tidal Disruption Accretion}

\author{Seungjae Lee}
\affiliation{Center for Theoretical Physics, Department of Physics and Astronomy, Seoul National University, Seoul 08826, Korea}

\author[0000-0003-4464-1160]{Ji-hoon Kim}
\correspondingauthor{Ji-hoon Kim, \href{mailto:me@jihoonkim.org}{me@jihoonkim.org}}
%\email{me@jihoonkim.org}
\affiliation{Center for Theoretical Physics, Department of Physics and Astronomy, Seoul National University, Seoul 08826, Korea}
\affiliation{Seoul National University Astronomy Research Center, Seoul 08826, Korea}

\author{Boon Kiat Oh}
\affiliation{Center for Theoretical Physics, Department of Physics and Astronomy, Seoul National University, Seoul 08826, Korea}

\begin{abstract}

Stars that are tidally disrupted by the massive black hole (MBH) may contribute significantly to the growth of the MBH, especially in dense nuclear star clusters (NSCs). 
Yet, this tidal disruption accretion (TDA) of stars onto the MBH has largely been overlooked compared to the gas accretion (GA) channel in most numerical experiments until now. 
In this work, we implement a black hole growth channel via TDA in the high-resolution adaptive mesh refinement code {\sc Enzo} to investigate its influence on a MBH seed's early evolution.
We find that a MBH seed grows rapidly from $10^3\msun$ to $\gtrsim 10^6\,\mathrm{M}_\odot$ in 200\,Myrs in some of the tested simulations.
Compared to a MBH seed that grows only via GA, TDA can enhance the MBH's growth rate by up to more than an order of magnitude.
However, as predicted, TDA mainly helps the early growth of the MBH (from $10^{3-4}\msun$ to $\lesssim10^{5}\,\mathrm{M}_\odot$) while the later evolution is generally dominated by GA.
We also observe that the star formation near the MBH is suppressed when TDA is most active, sometimes with a visible cavity in gas (of size $\sim$ a few pc) created in the vicinity of the MBH. 
It is because the MBH may grow expeditiously with both GA and TDA, and the massive MBH could consume its neighboring gas faster than being replenished by gas inflows.  
Our study demonstrates the need to consider different channels of black hole accretion that may provide clues for the existence of supermassive black holes at high redshifts.

\end{abstract}


%%%%%%%%%%%%%%%%%%%%%%%%%%%%%%%%%%%%%%%%%%%%%%%%%%
%%%%%%%%%%%%%%%%% BODY OF PAPER %%%%%%%%%%%%%%%%%%

\section{Introduction}
\label{sec: intro}

Discovered at the centers of most massive galaxies are the massive black holes (MBHs) with masses $\gtrsim 10^6\msun$. 
While the physical size of a MBH is insignificant compared to that of its host galaxy, the dynamical influence of the MBH extends throughout the entire host galaxy. 
In our contemporary understanding, most of the MBH mass is believed to have been acquired from the  accreting gas \citep{Bondi1952MNRAS.112..195B,Salpeter1955ApJ...121..161S, Salpeter1964ApJ...140..796S,Zeldovich1964SPhD....9..195Z, Lynden1969Natur.223..690L,Lynden1971MNRAS.152..461L,Lynden1978PhyS...17..185L}. 
However, the radiation from the accretion disk limits the gas accretion rate (GAR) to the so-called Eddington rate, making it challenging to explain the observed massive quasars at high redshifts \citep[for reviews, see][]{Inayoshi2020ARA&A..58...27I}. 
Thus, many studies have been conducted to understand the formation and growth history of MBHs \citep[for reviews, see][]{Volonteri2010A&ARv..18..279V,Sesana2012AdAst2012E..12S,Kormendy2013ARA&A..51..511K}.

In hydrodynamics simulations that probe the formation and evolution of galaxies, MBH physics has been a crucial component. 
Ever since the first implementation of MBHs in a galaxy-scale simulation by \cite{Springel2005MNRAS.361..776S}, simulators have shown that the energy from an accreting MBH is essential in preventing the overcooling of gas \citep[e.g.,][]{Sijacki2009MNRAS.400..100S, Booth2009MNRAS.398...53B, Dubois2010MNRAS.409..985D}. 
They also have demonstrated that the MBHs' evolution is closely linked to the star formation of their host galaxies \citep[e.g.,][]{Springel2005MNRAS.361..776S, Springel2005ApJ...620L..79S, Di2005Natur.433..604D, Hopkins2006ApJ...652..864H, Johansson2009ApJ...707L.184J, Kim2011ApJ...738...54K, 2019ApJ...887..120K, Choi2014MNRAS.442..440C}. 
The growth mechanism of the MBH itself has also been a topic of great interest. 
For example, how the infalling gas overcomes the angular momentum barrier has been discussed by many authors \citep[e.g.][]{Hopkins2010MNRAS.407.1529H, Hopkins2011MNRAS.415.1027H, Emsellem2015MNRAS.446.2468E}. 
However, MBH physics in a galaxy-scale hydrodynamics simulation is still far from complete. 
First and foremost, in most previous numerical studies, MBHs grow only via the gas accretion (GA) channel. 
And these GA models are often resolution-dependent \citep{Booth2009MNRAS.398...53B} and also rely on the assumption of spherically symmetric gas inflows in many implementations. 

Meanwhile, in the vicinity of a MBH seed at the center of a galaxy, a significant fraction of mass exists in the form of stars, not just in gas. 
When a star approaches a compact object such as a BH closer than its tidal radius, the star is disrupted by the compact object's tidal force, and a large fraction of the resulting stellar debris eventually accretes to the compact object \citep{Rees1988Natur.333..523R, Strubbe2009MNRAS.400.2070S}. 
A sufficient number of tidal disruption events (TDEs) could help the BH to grow more rapidly than was previously thought.

This may be especially true for the MBHs residing in the very dense nuclear star clusters (NSCs)  often found in the centers of galaxies.
The typical masses of NSCs are in the range $10^4\msun <  M_\mathrm{NSC} < 10^9\msun$ while their central stellar densities may exceed $10^6\,\mathrm{M}_\odot \mathrm{pc}^{-3}$   \citep[for reviews, see e.g.,][]{Neumayer2020A&ARv..28....4N}. 
In addition, the NSCs are found to coexist with MBHs in many galaxies \citep[e.g.,][]{Lauer1998AJ....116.2263L, Schodel2018A&A...609A..27S}, and the correlations between NSCs and MBHs have been widely discussed \citep[e.g.,][]{Seth2008ApJ...678..116S, Graham2009MNRAS.397.2148G, Nguyen2018ApJ...858..118N}.
Some authors have suggested that dense star clusters such as NSCs could be the birthplaces of intermediate-mass black holes (IMBHs; $10^2\msun < M_\mathrm{BH} < 10^5\msun$). 
The gravothermal contraction and the ensuing core collapse significantly increase the central density of a star cluster  \citep{Aarseth1974A&A....37..183A, Giersz1994MNRAS.269..241G, Takahashi1995PASJ...47..561T}, and then, runaway collisions between stars inside this dense core could lead to the formation of an MBH seed of mass $\gtrsim 10^2\msun$  \citep{Begelman1978MNRAS.185..847B, Quinlan1987ApJ...321..199Q, Ebisuzaki2001ApJ...562L..19E, Zwart2002ApJ...576..899P, Gurkan2004ApJ...604..632G, Zwart2004Natur.428..724P, FreitageII2006MNRAS.368..141F, FreitagI2006MNRAS.368..121F, Miller2012ApJ...755...81M}.
Even after the MBH seed has formed, there are still numerous stars in its vicinity inside the NSC. 
While the star-star collision must be nearly head-on for them to merge, the now massive MBH seed may tidally disrupt stars even when they are not in a head-on collision course. 
For example, combining observations and theories, \cite{Wang2004ApJ...600..149W} predicted that the TDEs are prevalent near a MBH, and the tidal disruption rate (TDR) varies inversely with the MBH mass.
Although it is rather uncertain whether we can generalize their result to the MBHs of masses $\lesssim 10^5\msun$, their work showed that a young MBH may have grown via successive TDEs. 
Using an analytical approach, \cite{Stone2017MNRAS.467.4180S} also argued that MBHs can form and grow inside NSCs.

Despite its importance, no numerical study has considered the tidal disruption accretion (TDA) channel of MBH growth in a galaxy-scale (or star cluster-scale) hydrodynamics simulation.  
One of the reasons is that the scattering between a star and a MBH is not properly resolved in a typical galaxy-scale hydrodynamics simulation --- unless the spatial resolution is vastly increased near the MBH.  
Indeed, given the current computational constraints, stellar scatterings or TDEs can be explicitly resolved only with a direct summation code such as $N${\sc-body6++gpu} \citep{Wang2015MNRAS.450.4070W}. 
Yet, these direct $N$-body codes tend not to include the hydrodynamics solver required to describe GA. 
Therefore, a sub-resolution prescription for TDA in a high-resolution hydrodynamic simulation can be the first step towards bridging the two different numerical approaches. 

Recently \cite{Pfister2021MNRAS.500.3944P} devised a TDA model in the {\sc Ramses} code and studied the rate of TDEs in galaxy mergers.
However, their work did not fully investigate the MBH evolution itself.  
In addition, their TDA model computes the TDR based on the stellar profile found in the simulation. 
Because a simulation with insufficient resolution may not follow the scattering process between stars and MBHs precisely, the stellar profile in the simulation may give inaccurate TDR estimates (for more discussion, see Section \ref{subsec: limitations}). 
Therefore, in the present study, we implement a TDA model based on a different approach to mitigate the issue, while focusing on the growth of a MBH seed after its formation in a NSC.  

The paper is organized as follows. 
In Section \ref{sec: Tidal Disruption Accretion Model}, we present our TDA model: theoretical backgrounds, and how we implement the TDA model in the adaptive mesh refinement (AMR) hydrodynamics code {\sc Enzo}. 
In Section \ref{sec:Simulations}, we introduce the initial condition and the parameters of the simulations. 
The results of our galaxy--NSC--MBH co-evolution simulations are described in Section \ref{sec:Result}, and we discuss the limitations of our work and future studies in Section \ref{sec:Discussion}. 
In Section \ref{sec:Conclusion}, we conclude our paper with the key findings.

\vspace{2mm}

\section{Tidal Disruption Accretion Model}
\label{sec: Tidal Disruption Accretion Model}

\vspace{1mm}

\subsection{Methodology Overview}

If a star approaches a compact object such as a BH, a MBH, or a neutron star closer than the tidal radius, the star is disrupted due to the tidal force of the compact object. 
For a star with mass $m_\star$ and radius $r_\star$ approaching a MBH of mass $M_\mathrm{BH}$, we can estimate the tidal radius $R_\mathrm{T}$ of the MBH as
\begin{align}
    R_\mathrm{T} &= r_\star\left(\frac{\eta^2M_\mathrm{BH}}{m_\star}\right)^{\frac{1}{3}} \nonumber \\  
    &\simeq \,2.25\times10^{-8}\eta^{\frac{2}{3}}\left(\frac{r_\star}{{\rm R}_\odot}\right)\left( \frac{M_\mathrm{BH}}{m_\star}\right)^{\frac{1}{3}} \mathrm{pc},
\label{eq: tidal radius}
\end{align}
where $\eta$ is the Safronov number that depends on the inner structure of the star, and ${\rm R}_\odot$ is the solar radius. 
After the TDE, a fraction $f$ of the resulting stellar debris accretes to the MBH ($f\lesssim1.0$), while the rest escapes the MBH's gravity \citep{Frank1976MNRAS.176..633F, Strubbe2009MNRAS.400.2070S}.  

In an ideal numerical simulation, one might imagine simply removing a star particle that encounters the MBH particle within $R_\mathrm{T}$ and add the star's mass to the MBH's. 
However, the length scale of $R_\mathrm{T}$ --- and the scattering process within --- is too small to be resolved in a typical galaxy-scale simulation that depicts the co-evolution of a MBH and its host galaxy. 
Hence, we have adopted a method that computes the TDR based on a statistical approach, also known as the {\it loss cone theory} \citep{Frank1976MNRAS.176..633F, Syer1999MNRAS.306...35S, Magorrian1999MNRAS.309..447M, Wang2004ApJ...600..149W}. 
The loss cone is the area in the phase space that is within or adjacent to the tidal radius of the MBH. 
If a stellar orbit lies in the loss cone, the star is thought to be a candidate for TD. 
Once we estimate the TDR, $\dot{N}_\mathrm{TD}$, we obtain the tidal disruption accretion rate (TDAR) of the MBH by
\begin{equation}
    \dot{M}_\mathrm{BH,\,TDA} = f \times m_{\star} \times \dot{N}_\mathrm{TD}
    \label{eq:mdot_BH}
\end{equation}
assuming that the stars have identical masses $m_\star$. 
The rest of this section describes how we estimate $\dot{N}_\mathrm{TD}$ in a simulation.

%\vspace{1mm}

\subsection{Key Assumptions}
\label{subsec:power-law}

In many observational studies, the stellar density profiles at the galactic centers are typically fitted by a power law $\rho_\star(r) \propto r^{-\gamma}$ with a power-law index $\gamma$ \citep[e.g.,][]{Faber1997AJ....114.1771F, Lauer1998AJ....116.2263L, Ferrarese2006ApJ...644L..21F, Ferrarese2006ApJS..164..334F, Schodel2009A&A...502...91S}. 
For most galaxies, $\gamma$ is between 1.0 and 2.0. 
Theoretically, this power law can be understood as a feature of a self-consistent stellar system \citep{binney2011galactic}. 
For example, a self-gravitating system following the Maxwell-Boltzmann distribution with a constant velocity dispersion gives us $\gamma = 2.0$. 
The outer region of a collapsing single-mass Plummer model yields $\gamma = 2.23$ \citep{Takahashi1995PASJ...47..561T}. 
Also, a stellar system dominated by a single MBH follows the power-law profile with $\gamma = 1.75$ \citep{Bahcall1976ApJ...209..214B}. 

Based on these observations, when estimating the TDR we assume that the NSC follows a power-law stellar density profile. 
However, unfortunately, it is not desirable to completely trust the density profile in a typical galaxy-scale simulation near the MBH since the stellar distribution there is under-resolved. 
Therefore, we use a sub-resolution, analytic approach to infer the stellar profile at the NSC's center needed to compute the TDR.
A naive, simple power-law profile such as the one assumed in \cite{Wang2004ApJ...600..149W} may work well for  high-mass MBHs ($M_\mathrm{BH} \gtrsim 10^5\,\msun$), but may over-estimate the TDR for low-mass MBHs ($M_\mathrm{BH} \lesssim 10^4\,\msun$).\footnote{The \cite{Wang2004ApJ...600..149W} model predicts that, for a fixed power-law stellar density profile with $\gamma > 1.42$ (see Eq.(\ref{eq: r_crit}) and the description below), the TDR varies inversely with the MBH mass.  Applying their model to the lower BH mass range  ($M_\mathrm{BH} \lesssim 10^4\,\msun$) returns an exceedingly  high TDR ($\dot{N}_\mathrm{TD} \gg 10^{-2}\,\mathrm{yr}^{-1}$), which is a sign that the TDR could be over-estimated with a naive assumption for the stellar profile.}
Thus, instead, while we use the simple power-law stellar density profile for high-mass MBHs, we consider a ``cored'' power law for low-mass MBHs, as motivated by \cite{Stone2015ApJ...806L..28S}: 
\begin{equation}
    \rho_\star\left(r\right) = \rho_\mathrm{c}\left(1 + {r^2}/{r_\mathrm{c}^2}\right)^{-{\gamma \over 2}},
\label{eq: power-law with core}
\end{equation}
where $\rho_\mathrm{c}$ and $r_\mathrm{c}$ are the density and size of the core, respectively.\footnote{In practice, our model entails that a low-mass MBH sits in a uniform density ($\rho_\mathrm{c}$) stellar core, while a high-mass MBH is at the center of a simple power-law profile.} 
Two more assumptions simplify our model:  
{\it (i)} the masses of stars are identically $m_\star$ (value to be chosen by the user; see Eq.(\ref{eq: deltaM_BH})), and
{\it (ii)} the MBH stays at the center of a NSC without ``wandering'' (for discussion, see Appendix \ref{subsec: constraining mbh seed mass}).

\vspace{1mm}

\subsection{Tidal Disruption Rate (TDR) Estimation}
\label{subsec:How to compute TDR}

We now discuss how the TDR is calculated using the stellar density profile assumed in  Section \ref{subsec:power-law}, and the various stellar distribution properties found in a simulation. 
In our prescription that computes the TDR, we first estimate the average stellar density $\rho_0$ and the average 1D stellar velocity dispersion $\sigma_0$ in a simulation inside a sphere of a user-defined radius $R_\mathrm{s}$ from the MBH ($R_\mathrm{s}$ is  set to $0.5\,\mathrm{pc}$, a value close to the minimum cell size $\Delta x_\mathrm{min}$ in a simulation; see Section \ref{subsec:refinement}). 
$r_0$ is then defined as the radius from the MBH at which the power-law density profile and the velocity dispersion profile becomes $\rho_0$ and $\sigma_0$, respectively.  
Also, the power-law index $\gamma$ and the size of the stellar core, $r_\mathrm{c}$, are manually assumed by the user (see Eq.(\ref{eq: power-law with core}) and Table \ref{table: runs}), but not the core density $\rho_\mathrm{c}$ which will later be estimated by our TDA model.

\begin{figure*}
    \vspace{-1mm}
    \hspace*{1mm}     
    \includegraphics[width=0.96\textwidth]{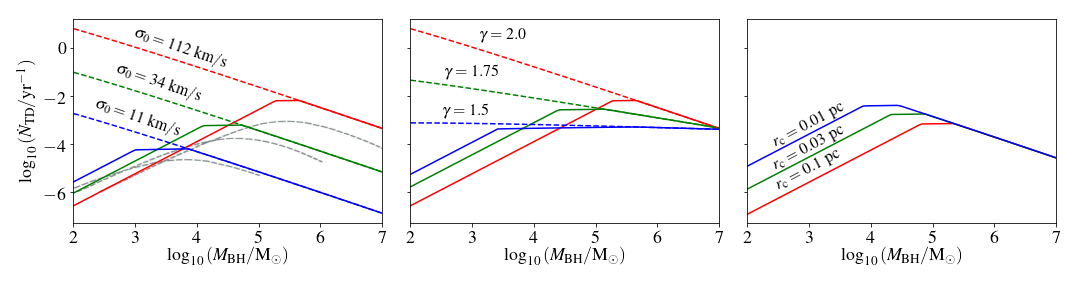}    
    \vspace{-4mm}
    \caption{Tidal disruption rate (TDR) as a function of massive black hole (MBH) mass. 
    The {\it solid lines} indicate the TDR of our model (see Section \ref{subsec:How to compute TDR}) while the {\it dashed lines} are the estimates by \citet[][i.e., no core; $r_\mathrm{c} = 0$]{Wang2004ApJ...600..149W}.  
    The {\it left panel} shows how the TDR varies with $\sigma_0$ near MBH with $\gamma = 2.0$:  $\sigma_\mathrm{0} = \mathrm{112}\,\mathrm{km\,s^{-1}}$ ($r_\mathrm{c} = 0.1\,\mathrm{pc}$), $\sigma_\mathrm{0} = 34\,\mathrm{km\,s^{-1}}$ ($r_\mathrm{c}=0.03\,\mathrm{pc}$), and $\sigma_\mathrm{0} = 11\,\mathrm{km\,s^{-1}}$ ($r_\mathrm{c}=0.01\,\mathrm{pc}$). 
    The $r_\mathrm{c}$ values are chosen in tandem with $\sigma_0$ to match the estimates in Figure 8 of \cite{Stone2017MNRAS.467.4180S} shown in this panel as {\it gray dashed lines}.
    The {\it middle panel} depicts the TDR for different $\gamma =$ 1.5, 1.75, 2.0 with $r_\mathrm{c} =$ 0.1 pc and $\sigma_\mathrm{0} = 112\,\mathrm{km\,s^{-1}}$. 
    Lastly, the {\it right panel} shows the TDR for different $r_\mathrm{c} =$ 0.01, 0.3, 0.1 pc with $\gamma = 2.0$ and $\sigma_\mathrm{0} = 112\, \mathrm{km\,s^{-1}}$.}
    \label{fig:Expectations}
    \vspace{1mm}
\end{figure*}

Then, the 1D velocity dispersion profile is written for a spherically symmetric system as 
\begin{align}
    \sigma^2\left(r\right) &= \frac{G}{\rho_\star\left(r\right)}\int^{\infty}_r \frac{M\left(r'\right)\rho_\star\left(r'\right)}{r'^2}dr'  \\
    &= \frac{2\pi G \rho_0 r^2_0}{\left(3 - \gamma\right)\left(\gamma - 1\right)}\left(\frac{r}{r_0}\right)^{2-\gamma}, 
\label{eq: sigma}
\end{align}
where we used  $\rho_\star\left(r\right) \simeq \rho_0\left(r/r_0\right)^{-\gamma}$, an approximation of Eq.(\ref{eq: power-law with core}) to simplify the integration \citep{binney2011galactic}.\footnote{This approximation does not make a significant difference in the obtained $\sigma(r)$ or $\sigma_\mathrm{c}$ because the integration is dominated by $r > r_\mathrm{c}$.} 
Plugging $r=r_0$ into Eq.(\ref{eq: sigma}) yields 
\begin{equation}
    \sigma_0^2 = \frac{2\pi G \rho_0 r^2_0}{\left(3 - \gamma\right)\left(\gamma - 1\right)}, 
\end{equation}
from which, as expected, $r_0$ is determined by $\rho_0$ and $\sigma_0$ as
\begin{align}
    r_0 &= \sqrt{\frac{\left(3-\gamma\right)\left(\gamma - 1\right)}{2\pi} \frac{\sigma_0^2}{G \rho_0}}  \nonumber \\
         &\simeq 2.0\,\,\mathrm{pc}\times\left(\frac{\sigma_0}{100\,\,\mathrm{km}\,\mathrm{s}^{-1}}\right)\left(\frac{\rho_0}{10^5\,\mathrm{M_\odot}\,\mathrm{pc}^{-3}}\right)^{-{1 \over 2}}.
\label{eq: r_0}
\end{align}
In addition, plugging $r=r_\mathrm{c}$ into Eq.(\ref{eq: sigma}) gives the 1D stellar velocity dispersion in the core as 
\begin{equation}
    \sigma_\mathrm{c}^2 = \frac{2 \pi G \rho_0 r_0^2}{\left(3 - \gamma\right)\left(\gamma - 1\right)}\left(\frac{r_\mathrm{c}}{r_0}\right)^{2-\gamma}, 
\end{equation}
whereas, from Eq.(\ref{eq: power-law with core}), the core stellar density is   
\begin{align}
    \rho_\mathrm{c} = \rho_\mathrm{0}(1+r_0^2/r_\mathrm{c}^2)^{\gamma \over 2}  \,\, \simeq \,\,\rho_0 \left(r_\mathrm{0}/r_\mathrm{c}\right)^{\gamma},
\label{eq: rho_c}
\end{align}
where $r_\mathrm{0}^2/r_\mathrm{c}^2 \gg 1$ because $ r_0 \sim 1 \,\mathrm{pc}$ and $r_\mathrm{c} \sim [0.03, 0.3]\,\mathrm{pc}$ are assumed in our estimation (see Table \ref{table: runs} and Appendix \ref{subsec: constraining TDA parameters}).  

\begin{table*}
\vspace{0mm}
\caption{Structural characteristics of a model galaxy in our fiducial initial condition}
\vspace{-3mm}
\centering
\begin{tabular}{l|lll}
\hline
\hline
 & Density profile & Structural properties\tablenotemark{\scriptsize \textcolor{red}{\textdagger}} & Data type  \\[0.5mm]
\hline
\hline
 Dark matter halo & \citet{Navarro1997ApJ...490..493N} \,\,\,& $M_{200} = 1.074\times 10^{12}\msun$, \,\,$v_\mathrm{c,\,200} = 150\,\km\s^{-1}$, \,\,\,& $10^7$ particles \\
  & & $r_{200} = 205.5\,\kpc$, \,\,$c = 10$, \,\,$\lambda = 0.04$ & $(m_\mathrm{DM} = 1.791\times 10^5 \msun)$ \\[0.5mm]
\hline
 Stellar disk & exponential &  $M_{\mathrm{d, \,star}} = 3.292\times 10^{10}\msun$, & $10^7$ particles \\
  & & $r_{\mathrm{d,\, star}} = 3.432\,\kpc$, $\,\,{z_\mathrm{\,d,\, star}} = 0.1 r_{\mathrm{d,\, star}} = 343\,\pc$ & $(m_{\mathrm{d}} = 3.292\times 10^3 \msun)$ \\[0.5mm]
\hline  
 Gas disk & exponential & $M_\mathrm{d, \, gas} = 8.593 \times 10^{9}\msun$, & adaptive mesh\\
  & & $r_\mathrm{d, \, gas} = r_{\mathrm{d,\, star}}  = 3.432\,\kpc$, $\,\,z_\mathrm{\,d,\,gas} = z_\mathrm{d,\,star} = 343\,\pc$ \,\,& \\[0.5mm]  
\hline  
 Stellar bulge & \citet{Dehnen1993MNRAS.265..250D} & $M_\mathrm{b} = 4.115\times 10^9\msun$, & $1.25\times 10^6$ particles\\
  & & $r_\mathrm{\,b} = 0.1 r_\mathrm{d,\, star} = 343\, \pc$, \,\,$\gamma_\mathrm{\,b} = 2.0$, \,\,$r_\mathrm{cutoff,\,b} = 50 \pc$ \,\,\, & $(m_\mathrm{b} = 3.292\times 10^3\msun)$ \\[0.5mm]
\hline  
 Nuclear star cluster (NSC) & \citet{Dehnen1993MNRAS.265..250D} &  $M_\mathrm{NSC} = 10^7\msun$,  & $2\times 10^4$ particles\\
  & & $r_\mathrm{\,NSC} = 3.0\pc$, \,\,$\gamma_\mathrm{\,\,NSC} = 2.0$, \, $r_\mathrm{cutoff,\,NSC} = 0.5\pc$  & $(m_\mathrm{NSC} = 500\,\msun)$ \\[0.5mm]
\hline  
 Massive black hole (MBH) & N/A &  $M_\mathrm{BH, \, init} = 10^3 \msun\,\,\,\,\,\, ({\rm or}\,\,\, 8 \times 10^3 \msun )$ & a single particle \\[1mm]  
\hline
\end{tabular}
\vspace{-1mm}
\tablecomments{\scriptsize \,\,\,{\textdagger}\,For the detailed explanations on these parameters adopted for our suite of galaxy--NSC--MBH co-evolution simulations, see Section \ref{subsec:IC}.}
\label{table: Initial Conditions}
\vspace{0mm}
\end{table*}

Now we consider two cases:  
\begin{itemize}
\item If $M_\mathrm{BH}$ is small (i.e., $M_\mathrm{BH} < M_\mathrm{c}$; see Eq.(\ref{eq: TDE 1})), the gravitational influence of the MBH is restricted to the uniform stellar core. 
In such a system, we adopt the TDR estimate in \cite{Rees1988Natur.333..523R}:
\begin{IEEEeqnarray}{rCl}
\dot{N}_\mathrm{Rees} & \simeq &\ 10^{-4}\,\mathrm{yr}^{-1} \left(\frac{M_\mathrm{BH}}{10^6\, \mathrm{M}_\odot}\right)^{4 \over 3} \nonumber
\\
&& \times \left(\frac{n_\mathrm{c}}{10^5\,\mathrm{pc}^{-3}}\right) \left(\frac{\sigma_\mathrm{c}}{100\,\,\mathrm{km}\,\mathrm{s}^{-1}}\right)^{-1},
\label{eq: Rees1988}
\end{IEEEeqnarray}
where $n_\mathrm{c} = \rho_\mathrm{c}/m_\star$ is the stellar number density at the core (the value of $m_\star$ to be chosen by the user; see Eq.(\ref{eq: deltaM_BH})). 
Here,  $\dot{N}_\mathrm{Rees}$ increases with $M_\mathrm{BH}$.

\item If $M_\mathrm{BH}$ is large (i.e., $M_\mathrm{BH} > M_\mathrm{c}$; see Eq.(\ref{eq: TDE 1})), the gravitational influence of the MBH reaches beyond the core radius. 
As stated in Section \ref{subsec:power-law} we assume that the stellar density follows a simple power law. 
Then, according to \cite{Wang2004ApJ...600..149W}, the TDR becomes
\begin{align}
\dot{N}_\mathrm{WM} &\approx \frac{M_\star\left(r_\mathrm{crit}\right)}{m_\star\, t_\mathrm{\,R}\left(r_\mathrm{crit}\right)} \\
        &\simeq \left(3-\gamma\right)(\ln{\Lambda})\, G^{1\over 2}\rho_0^2 r_0^{9 \over 2}M_\mathrm{BH}^{-{3 \over 2}}\left(\frac{r_\mathrm{crit}}{r_0}\right)^{{9\over 2}-2\gamma},
\label{eq: TDE WM}
\end{align}
where $M_\star(r)$ is the stellar mass enclosed in $r$, $t_\mathrm{\,R}(r)$ is the stellar relaxation time at radius $r$, and $\ln \Lambda =  \ln (0.4\, {M_\mathrm{BH}}/ {m_\star})$ is the Coulomb logarithm \citep{Spitzer1958ApJ...127..544S}.\footnote{Here in Eq.(\ref{eq: TDE WM}), the deformation of stellar distribution by the central MBH \citep{Bahcall1976ApJ...209..214B}  is not considered.}  
$r_\mathrm{crit}$ is the radius at which a star is scattered through an angular size of the loss cone in its dynamical time and is related to $\rho_0$ and $r_0$ as
\begin{IEEEeqnarray}{rCl}
\left(\frac{r_\mathrm{crit}}{r_0}\right)^{4-\gamma} & = & \frac{\eta^{2 \over 3}}{\pi}\left(\frac{r_\star^3}{m_\star}\right)^{1\over 3}m_\star^{-1}\nonumber
\\
&& \negmedspace {} \times \rho_0^{-1}r_0^{-4}\left(\ln{\Lambda}\right)^{-1}M_\mathrm{BH}^{7\over 3}\,,
\label{eq: r_crit}
\end{IEEEeqnarray}
for which we set $\eta = 0.844$ and $r_\star= {\rm R}_{\odot}\left({m_\star}/{{\rm M}_{\odot}}\right)^{1/3}$ for our simulations.\footnote{Two angles appear in the loss cone theory. First, the ``angular size'' of the loss cone at radius $r$ is defined as $\theta_\mathrm{\,lc}(r) = ({R_\mathrm{T}} / {r^2}) (GM_\mathrm{BH}/{\sigma}(r))$. Recall that $R_\mathrm{T}$ is the tidal radius defined in Eq.(\ref{eq: tidal radius}). Second, $\theta_\mathrm{d}(r)$ is the angle through which a star is scattered in a dynamical time $t_\mathrm{d}(r) = r/\sigma(r)$. If $\theta_\mathrm{\,lc} > \theta_\mathrm{d}$, the loss cone is in the diffusive regime and becomes empty. If $\theta_\mathrm{\,lc} < \theta_\mathrm{d}$, the loss cone is always filled with stars. The radius at which these two angles are equal is called the critical radius, $r_\mathrm{crit}$.}$^{,}$\footnote{Our estimate weakly depends on the choice of $r_\star$. For example, inserting $r_\star= {\rm R}_{\odot}\left({m_\star}/{{\rm M}_{\odot}}\right)^{0.8}$ gives a rate decreased by $\lesssim 5$\%.}
Here, $\dot{N}_\mathrm{WM}$ decreases with $M_\mathrm{BH}$ for $\gamma > 1.42$ because $t_\mathrm{R}$ increases with $M_\mathrm{BH}$ (i.e., refilling the loss cone becomes harder as the MBH grows). 
\end{itemize}

From these considerations we build a model for the TDR by combining Eqs.(\ref{eq: Rees1988}) and (\ref{eq: TDE WM}):
\begin{equation}
    \dot{N}_\mathrm{TD,1}\left(M_\mathrm{BH}, \rho_0, \sigma_0, \gamma, r_{\rm c} \right) = 
    \begin{cases}
        \dot{N}_\mathrm{Rees}  & \text{if}\ M_\mathrm{BH} < M_\mathrm{c}, \\
        \dot{N}_\mathrm{WM}  & \text{if}\ M_\mathrm{BH} > M_\mathrm{c},
    \end{cases}
\label{eq: TDE 1}
\end{equation}
where the transition mass $M_\mathrm{c} \,\,(\sim 10^{4-5} \,\mathrm{M}_\odot) $ is the mass of a MBH that makes $r_\mathrm{crit}(M_\mathrm{BH})$ equal to $r_\mathrm{c}$.\footnote{Note that Eq.(\ref{eq: r_crit}) tells us that $r_\mathrm{crit}$ is a function of $M_\mathrm{BH}$.  Because the Eq.(\ref{eq: TDE WM}) assumes that the MBH is embedded in a power-law profile, it requires $r_\mathrm{crit} \gtrsim r_\mathrm{c}$. Therefore, in principle, the transition between the two estimates, Eqs.(\ref{eq: Rees1988}) and (\ref{eq: TDE WM}), should occur at $r_\mathrm{crit} = r_\mathrm{c}$.} 
However, since the two estimates, Eqs.(\ref{eq: Rees1988}) and (\ref{eq: TDE WM}), do not exactly agree at $M_\mathrm{BH} = M_\mathrm{c}$, we limit Eq.(\ref{eq: TDE 1}) as  
\begin{equation}
    \dot{N}_\mathrm{TD}  \,\,= \,\,{\tt min} \left\{ \dot{N}_\mathrm{TD,1}, \,\,\dot{N}_\mathrm{Rees}({\scriptstyle \,M_\mathrm{BH}=M_\mathrm{c}}),  \,\,\dot{N}_\mathrm{WM}({\scriptstyle M_\mathrm{BH}=M_\mathrm{c}}) \right\}
\label{eq: Ndot_TDE}
\end{equation}
to remove the discontinuity at $M_\mathrm{c}$. By taking the minimum, we can also be conservative in our TDR estimates.

In Figure \ref{fig:Expectations}, we plot the resulting $\dot{N}_\mathrm{TD}\left(M_\mathrm{BH}\right) $ in order to observe its dependence on various parameters.  
First, for low-mass MBHs ($M_\mathrm{BH} \lesssim 10^4\,\msun$), our model predicts that $\dot{N}_\mathrm{TD}$ increases with $M_\mathrm{BH}$ (solid lines in {\it all} panels), while the extrapolation of the \cite{Wang2004ApJ...600..149W} model with no stellar core expects a decreasing $\dot{N}_\mathrm{TD}$ with respect to $M_\mathrm{BH}$ (dashed lines in the {\it left} and {\it middle} panels).
For high-mass MBHs ($M_\mathrm{BH} \gtrsim 10^5\,\msun$) in the {\it left} panel, the larger $\sigma_0$ is,  the higher $\rho_\mathrm{c}$ becomes, and so does $\dot{N}_\mathrm{TD}$.
Overall, this panel shows that our model behaves similarly as \citet[][see their Figure 8]{Stone2017MNRAS.467.4180S} seen here as gray dashed lines. 
In the {\it middle} panel, readers can notice that the slope index of the stellar profile, $\gamma$, determines the slope of $\dot{N}_\mathrm{TD}$ for high-mass MBHs ($M_\mathrm{BH} \gtrsim 10^5\,\msun$). 
The larger $\gamma$ is,  the higher $\rho_\mathrm{c}$ becomes for a given $\sigma_0$, and so does the peak value of $\dot{N}_\mathrm{TD}$. 
Lastly, in the {\it right} panel of Figure \ref{fig:Expectations}, varying $r_\mathrm{c}$ changes the peak value of $\dot{N}_\mathrm{TD}$ and the transition mass $M_\mathrm{c}$ between the two MBH mass regimes, Eqs.(\ref{eq: Rees1988}) and (\ref{eq: TDE WM}).  
But it does not change the slope of $\dot{N}_\mathrm{TD}$ for the high-mass MBHs because, for the chosen $\gamma = 2.0$, Eq.(\ref{eq: sigma}) predicts a constant velocity dispersion. 

%\vspace{1mm}
\subsection{Implementing the Tidal Disruption Accretion (TDA) Channel in the Simulation Code}
\label{subsec:TDA implementation}

Once the stellar density $\rho_0$ and dispersion $\sigma_0$ around the MBH are specified, we find the TDR with Eq.(\ref{eq: Ndot_TDE}). 
Then, from Eq.(\ref{eq:mdot_BH}) we increase the MBH mass at each timestep by
\begin{equation}
    \Delta M_\mathrm{BH,\,TDA} = f m_\star \,\dot{N}_\mathrm{TD}\,\Delta\, t
\label{eq: deltaM_BH}
\end{equation}
where $\Delta \,t$ is the size of the timestep. 
The mass of a disrupted star in our model, $m_\star$, is set to $0.7\,\mathrm{M}_\odot$, the average value of the Salpeter IMF \citep{Salpeter1955ApJ...121..161S}.
However, it should be noted that,  due to limited resolution, a newly-formed ``star particle'' in our simulation has  a mass of $m_{\rm star} \gtrsim 200\, M_\odot$.   
Therefore, in practice, we subtract $\Delta M_\mathrm{BH,\,TD}$ uniformly from all the ``star particles'' within $R_\mathrm{s}$ from the MBH ($R_\mathrm{s}=0.5\,\mathrm{pc}$; see Section \ref{subsec:How to compute TDR}), and then add it to the MBH. 
In other words, each star particle within $R_\mathrm{s}$ loses a fraction $\alpha$ of its mass  as  
\begin{equation}
    m_\mathrm{star} \,\, \rightarrow \,\,\,\, (1 - \alpha)\, m_\mathrm{star} = \left(1 - \frac{m_\star\, \dot{N}_\mathrm{TD}\,\Delta \,t}{M_\star(R_\mathrm{s})} \right)\, m_\mathrm{star}
\end{equation}
in which we set $f = 1.0$ in our simulations for simplicity.

\vspace{2mm}

\section{Simulations}
\label{sec:Simulations}

In this section, we present the initial conditions, refinement criteria, and baryonic physics employed in our suite of simulations. 
The high-resolution Eulerian AMR code {\sc Enzo} \citep{Bryan2014ApJS..211...19B, Brummel2019JOSS....4.1636B} provides the key zoom-in simulation technology and the baseline physics modules, and here we describe the ones that are closely related to the topic of our interest.  

\subsection{Initial Condition}
\label{subsec:IC}

For a fiducial initial condition, we adopt an isolated Milky Way-mass galaxy harboring a MBH and a NSC at its center.  
While we generate our initial condition using the code {\sc Dice} \citep{Perret2016ascl.soft07002P}, many galactic properties are motivated by one of the isolated galaxy initial conditions in the {\it AGORA} High-resolution Galaxy Simulations Comparison Project \citep{Kim2014ApJS..210...14K, 2016ApJ...833..202K, 2020arXiv200104354R, 2021ApJ...917...64R}.
The galaxy in our initial condition includes a dark matter halo, an exponential stellar/gas disk, a stellar bulge, a NSC, and a MBH seed (summarized in Table \ref{table: Initial Conditions}).

The dark matter halo follows the Navarro-Frank-White profile \citep[NFW;][]{Navarro1997ApJ...490..493N} with the circular velocity $v_\mathrm{c,\,200} = 150\,\,\mathrm{km}\,\mathrm{s}^{-1}$ and the virial mass $M_{200} = 1.072\times 10^{12}\msun$. 
The concentration parameter $c$ and the spin parameter $\lambda$ are set to 10 and 0.04, respectively. 
The dark matter halo is composed of $10^7$ particles with equal masses $m_\mathrm{DM} = 1.791\times 10^5\msun$. 
The masses of the stellar and gas disk are $M_{\mathrm{d,\, star}} = 3.292\times 10^{10}\msun$ and $M_\mathrm{d, \,gas} = 8.593\times 10^9\msun$, respectively. 
Each follows an exponential profile, e.g.,
\begin{equation}
    \rho_{\rm d,\,star}(r,z) = \rho_\mathrm{0, \,star} \,\exp({-{r}/{r_\mathrm{\,d, \,star}}})\exp({-{\left|z\right|}/{z_\mathrm{\,d, \,star}}})
\end{equation}
with the scale radius $r_\mathrm{d, \,star} = 3.432\,\mathrm{kpc}$, the scale height $z_\mathrm{d, \,star} = 0.1\,r_\mathrm{d, \,star}$, and $\rho_\mathrm{0,\,star} = {M_\mathrm{d, \,star}}/{(4\pi r_\mathrm{d, \,star}^2 \,z_\mathrm{d, \,star})}$. 
The stellar bulge of mass $M_\mathrm{b} = 4.297\times 10^9\msun$ follows the Dehnen profile \citep{Dehnen1993MNRAS.265..250D, Tremaine1994AJ....107..634T},
\begin{equation}
    \rho_\mathrm{b} \left(r\right) = \frac{\left(3-\gamma_\mathrm{\,b}\right)M_\mathrm{b}}{4\pi}\frac{r_\mathrm{\,b}}{r^{\gamma_\mathrm{\,b}} \left(r+r_\mathrm{\,b}\right)^{4-\gamma_\mathrm{\,b}}}
\end{equation}
with a half-mass radius $r_\mathrm{\,b} = 0.1\,r_\mathrm{d, \, star}$ and the index $\gamma_\mathrm{\,b}= 2.0$  \citep[can be regarded as a generalized version of][]{Hernquist1990ApJ...356..359H}.\footnote{In contrast, the initial condition in the {\it AGORA} Project \citep{2016ApJ...833..202K} follows the Hernquist profile which gives an infinite stellar density at $r = 0$.}  
In the initial condition, each ``star particle'' in the disk and bulge has a mass of  $3.292\times 10^3\msun$.

To set up a NSC with its own structural properties, separately from the stellar bulge, we cap the density of the central region of the bulge by setting a  ``cutoff'' radius, $r_\mathrm{cutoff,\,b} = 50\,\pc$ --- i.e., $\rho_\mathrm{b}(r<r_\mathrm{cutoff,\,b}) = \rho_\mathrm{b}(r_\mathrm{cutoff,\,b})$. 
Then, a NSC of mass $M_\mathrm{NSC} = 10^7 \mathrm{M}_\odot$ is inserted at the galaxy's center.
It follows the same Dehnen profile but with a smaller half-mass radius $r_\mathrm{\,NSC} = 3.0\,\pc$, a smaller cutoff radius $r_\mathrm{cutoff,\,NSC} = 0.5\,\pc$, and $\gamma_\mathrm{\,\,NSC} = 2.0$.\footnote{It should noted that the power-law profile with an index $\gamma_\mathrm{\,\,NSC} = 2.0$ is used to merely initialize the ``star particles'' within the NSC.  The particles' distribution later on may not follow the initial profile, and is likely unrealistic anyways because the central region of the NSC is under-resolved.  In our sub-resolution TDR estimates, we ``assume'' that the ``individual stars'' in the vicinity of the MBH follow the power-law profile, Eq.(\ref{eq: power-law with core}), with  $\gamma=1.5-2.0$ as listed in Table \ref{table: runs}.  See Section \ref{subsec: limitations} for more discussion.  \label{different_gammas}} 
Each ``star particle'' in the NSC represents $500\,\msun$, a value smaller than the initial MBH mass. 
Finally, we plant a MBH at the NSC's center --- with two choices of initial masses, $M_{\rm BH,\,init} = 10^3$ or $8\times 10^3\msun$ (for motivations for its value, see Appendix \ref{subsec: constraining mbh seed mass}). 
The MBH physics implemented is described in Sections \ref{subsec:TDA implementation} and \ref{subsec: MBH physics}.

\subsection{Refinement Strategy}
\label{subsec:refinement}

In the simulation box of $(1.307 \,\mathrm{Mpc})^3$, we apply AMR only in the pre-defined, innermost $(200\,\pc)^3$ box ($\sim$ size of the stellar bulge), dubbed {\tt RefineRegion}.  
On the outside, the {\tt RefineRegion} is surrounded by static nested volumes with successively coarser resolutions.  
The {\tt RefineRegion} is first uniformly resolved with cells of $5\pc$ width. 
Each of the cells is then adaptively split into $2^3$ child cells if the cell's stellar or gas mass exceeds a certain threshold which depends on the cell size $\Delta x$ (or the refinement level) as
\begin{equation}
    \begin{cases}
        \,\,M_\mathrm{ref, \,gas}^{\Delta x} = \sqrt{2^3 \left(\Delta x \over \,\,5\pc\,\, \right) } \times 5\times 10^2\msun\,, \\[2mm]
        \,\,M_\mathrm{ref, \, star}^{\Delta x} = \sqrt{2^3 \left(\Delta x \over \,\,5\pc\,\, \right) } \times 5\times10^4 \msun\,.
    \end{cases}
\end{equation}    
Therefore, the mass threshold for refinement decreases from $M_\mathrm{ref, \,gas}^{\Delta x=5\pc} = 2^{3/2} \times 5\times 10^2\msun$ at $\Delta x=5\pc$ down to $M_\mathrm{ref, \,gas}^{\Delta x=0.625\pc} = 5\times 10^2\msun$ at $\Delta x=0.625\,\pc=\Delta x_\mathrm{min} $ when the refinement stops. 
This is a super-Lagrangian refinement scheme to adaptively increase spatial resolution in the targeted region around the NSC at limited computational cost.  

\begin{figure}
    \hspace*{-3mm} 
    \vspace{2mm}
    \centering
    \includegraphics[width=1.02\columnwidth]{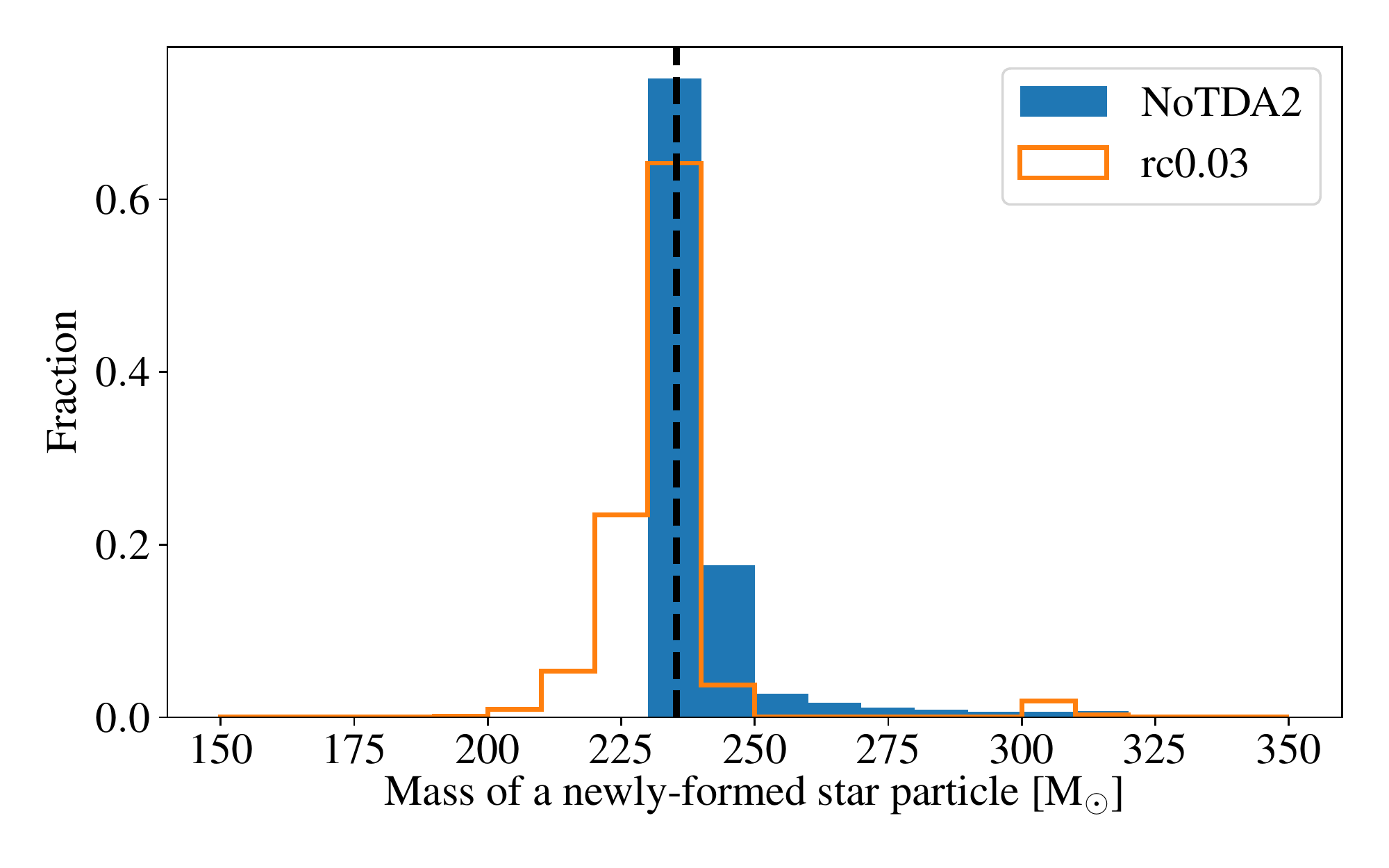}    
    \vspace{-9mm}
    \caption{The distribution of the masses of newly-formed ``star particles'' in our simulations. 
    The {\it black dashed line} is the lowest possible mass of a ``star particle'', $m_{\mathrm{star},\, \mathrm{min}} = 235\,\,\mathrm{M}_\odot$ set by the star particle formation criteria (after it loses a fraction of its mass via stellar feedback).  
    In a simulation without the tidal disruption accretion (TDA) of stars ({\it blue}), $m_{\mathrm{star},\, \mathrm{min}}$ is indeed the lowest mass in the distribution.  
    In contrast, in a simulation with TDA ({\it orange}), star particle masses can be below $m_{\mathrm{star},\, \mathrm{min}}$ as they may lose a fraction of their masses via TDA.   
    For detailed explanations, see Section \ref{subsec:sf}.} 
    \label{fig:Star mass distribution}
    \vspace{-1mm}
\end{figure}

\begin{table*}
\vspace{-1mm}
\caption{A suite of idealized galaxy--NSC--MBH co-evolution simulations listed with their runtime parameters}
\vspace{-3mm}
\centering
\begin{tabular}{c|c|cccc}
\hline
\hline
 & Run name & MBH seed mass  & Tidal disruption accretion (TDA)\tablenotemark{\scriptsize \textcolor{red}{\textdagger}} & NSC power-law index assumed  & NSC core size assumed \\[-0.8mm]
 & & $M_\mathrm{BH,\,init} \,\,\left[\mathrm{M}_\odot\right]$ & [on/off]& $\gamma$ & $r_\mathrm{c} \,\,\left[ \mathrm{pc} \right]$ \\[0.5mm] 
\hline
\hline
\multirow{4}{*}{\it{Set A}\tablenotemark{\scriptsize \textcolor{red}{\textdaggerdbl}}} & {\tt NoTDA1} & $8\times10^3$ & off \,(i.e., only gas accretion)& - & - \\
 & {\tt gamma1.5}  & $8\times10^3$ & on \,(i.e., gas accretion$+$TDA) & 1.5 & 0.1 \\ 
 & {\tt gamma1.75}  & $8\times10^3$ & on & 1.75 & 0.1 \\ 
 & {\tt gamma2.0} & $8\times10^3$ & on & 2.0 & 0.1 \\[0.2mm] 
 \hline
\multirow{5}{*}{\it{Set B}} & {\tt NoTDA2} & $10^3$ & off & - & - \\
 & {\tt rc0.03} & $10^3$ & on & 2.0 & 0.03 \\
 & {\tt rc0.05} & $10^3$ & on & 2.0 & 0.05 \\
 & {\tt rc0.1} & $10^3$ & on & 2.0 & 0.1 \\
 & {\tt rc0.3} & $10^3$ & on & 2.0 & 0.3 \\[0.2mm] 
\hline
\end{tabular}
\vspace{-1mm}
\tablecomments{\scriptsize \,\,\,{\textdagger}\,For the descriptions of our sub-resolution TDA model and its user-defined parameters such as the power-law index $\gamma$ of the NSC's density profile, and the size of its stellar core, $r_\mathrm{c}$, see Section \ref{subsec:How to compute TDR}. \,\,\,{\textdaggerdbl}\,``{\it Set A}'' is a suite of simulations with varying $\gamma$ while ``{\it Set B}'' includes the runs with varying $r_\mathrm{c}$.  For more information, see Section \ref{subsec:Initial Evolution of the simulations}.} 
\label{table: runs}
\vspace{0mm}
\end{table*}

\subsection{Star Particle Formation and Feedback}
\label{subsec:sf}

A ``star particle'' forms if the density of a maximally refined cell exceeds a certain threshold $\rho_\mathrm{th}$, and the cell cannot be reliably treated in the hydrodynamics solver. 
Our star particle formation model is based on \cite{Cen1992ApJ...399L.113C} and \cite{Kim2011ApJ...738...54K}. 
Following the Jeans argument, we choose the threshold $\rho_\mathrm{th} = { \pi c_\mathrm{s}^2}/{(G \lambda_\mathrm{J}^2)}$ with the Jeans length set to the finest spatial resolution, $\lambda_\mathrm{J} = \Delta x_\mathrm{min}$, and with the sound speed $c_\mathrm{s}$ at $100\,\mathrm{K}$. 
This gives $\rho_\mathrm{th} = 2571\,\,\mathrm{M}_\odot \mathrm{pc}^{-3}$ (or the threshold mass $M_\mathrm{J} = 628\,\,\mathrm{M}_\odot$ for the given $\Delta x_\mathrm{min} = 0.625\,\pc$). 
We turn 50\% of the cell's mass into a newly-formed ``star particle'' with its dynamical timescale set to 
$\tau_\mathrm{dyn} = 1\,\mathrm{Myr}$.
In the next $12\tau_\mathrm{dyn}$, the star particle contributes to the thermal supernovae feedback while returning 25\,\% of its mass back to gas phase. 
These considerations yield the lowest possible star particle mass of $m_\mathrm{star, \,min} = 235\,\,\mathrm{M}_\odot$.

Figure \ref{fig:Star mass distribution} shows the mass distribution of newly-formed star particles.  
In the baseline simulation without TDA ({\it blue} histogram for the {\tt NoTDA1} run; see Table \ref{table: runs} and Section \ref{subsec:TDA implementation}), $m_{\mathrm{star},\, \mathrm{min}}$ is indeed the lowest mass in the distribution.  
In contrast, in the run with TDA ({\it orange} histogram for the {\tt rc0.03} run), the masses of star particles can be lower than $m_{\mathrm{star},\, \mathrm{min}}$ because star particles near the MBH may lose their masses to the MBH via the TDA channel (see Section \ref{subsec:TDA implementation}).\footnote{Therefore, the present stellar masses may not be suitable when estimating the star formation rate (SFR). Instead, since the number of newly-formed star particles is preserved, we estimate the SFR by counting all the newly-formed star particles, and assuming that their masses at the time of their formation are equally $m_\mathrm{star, \,min} = 235\,\,\mathrm{M}_\odot$ (see Section \ref{subsec:results-sf}).} 

\subsection{Massive Black Hole (MBH) Accretion and Feedback}
\label{subsec: MBH physics}

The MBH particle inserted at the center of the simulation box (see Section \ref{subsec:IC}) grows via two channels:  GA and TDA.  
Then the total MBH accretion rate is simply
\begin{equation}
    \dot{M}_\mathrm{BH} = \dot{M}_\mathrm{BH, \,TDA} + \dot{M}_\mathrm{BH, \,GA}
\end{equation}
where the first term (TDAR) is from Eq.(\ref{eq:mdot_BH}) or (\ref{eq: deltaM_BH}) (see Section \ref{sec: Tidal Disruption Accretion Model}), and the second term (GAR) is from the conventional Bondi-Hoyle-Lyttleton formalism \citep{Bondi1944MNRAS.104..273B, Bondi1952MNRAS.112..195B} with the Eddington limit, 
\begin{align}
       \dot{M}_\mathrm{BH,\,GA} &= \min\left(\dot{M}_\mathrm{BH,\,Bondi},\,\, \dot{M}_\mathrm{BH,\,Edd}\right) \nonumber \\
       & = \min\left(\frac{4\pi G^2 M_\mathrm{BH}^2 \rho_\mathrm{B}}{c_\mathrm{s}^3} ,\,\, \frac{4\pi G M_\mathrm{BH}m_\mathrm{p}}{\epsilon_\mathrm{r}\sigma_\mathrm{T}c}\right),
\label{eq: mdot_gas}
\end{align}
for a MBH residing in a cell with the sound speed $c_\mathrm{s}$, where $m_\mathrm{p}$ is the proton mass, $\sigma_\mathrm{T}$ is the Thomson scattering cross section, and $\epsilon_\mathrm{r} = 0.1$ is the BH's radiative efficiency.  
$\rho_\mathrm{B} $ is the gas density at the Bondi radius $R_\mathrm{B} = {2G M_\mathrm{BH}}/{c_\mathrm{s}^2} $, and is estimated from the density $\rho_{\rm gas}$ of the cell where the MBH particle resides by $\rho_{\rm B} = \rho_{\rm gas} \,\cdot\, {\tt min} \{ (\Delta x/R_{\rm B})^{1.5}, 1.0 \}$ \citep[for details, see][]{Kim2011ApJ...738...54K, 2019ApJ...887..120K}.
The MBH particle returns thermal feedback energy to the cells around it at a rate of 
\begin{equation}
	L_\mathrm{BH} = \epsilon_\mathrm{r}\,\dot{M}_\mathrm{BH,\,GA} \cdot c^2
\end{equation}
which is proportional only to the GAR, $\dot{M}_\mathrm{BH,\,GA}$.

%\vspace{2mm}
\section{Results}
\label{sec:Result}

In this section, we analyze the effects of our TDA model on the growth of the MBH using the simulations with varying TDA parameters. 
We also examine the secondary effects arising from the TDA-boosted MBH accretion.

\begin{figure*}
%    \hspace*{-6mm} 
    \vspace{-1mm}
    \centering
    \includegraphics[width=0.86\textwidth]{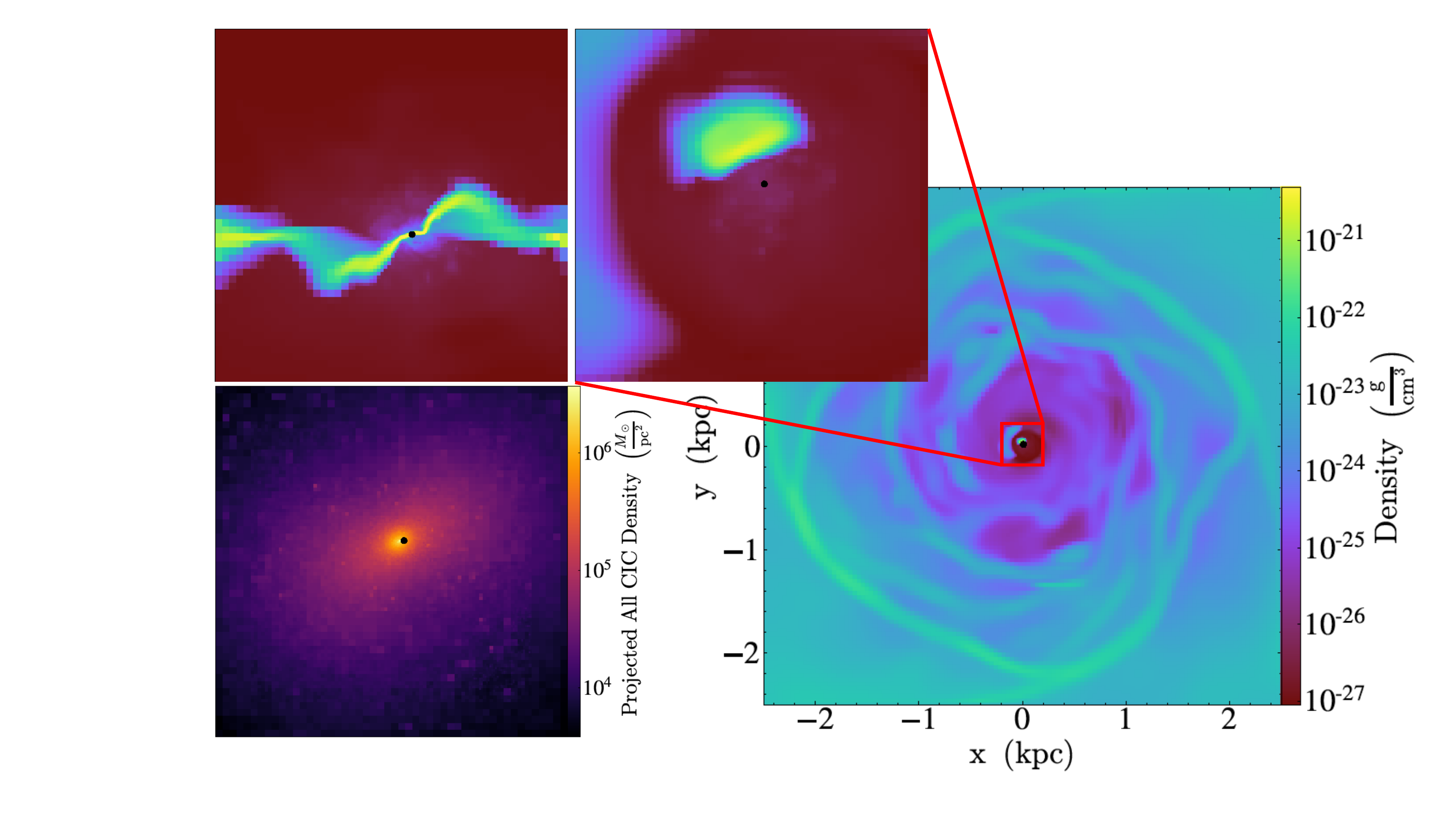}    
    \vspace{-5mm}
    \caption{Snapshots of the {\tt NoTDA2} run at $t$ = 50\,Myr. 
    The {\it right panel} is the face-on gas density of the central (5 kpc)$^2$ sliced through the location of the MBH. 
    Among the three zoomed-in images on the {\it left} (250\,pc width), the {\it top two panels} are the edge-on and face-on sliced gas density centered on the MBH, while the {\it bottom panel} is the projected stellar density. 
    The black dot in each image indicates the MBH's position.
    The figure illustrates the high resolution we retain near the MBH to apply the TDA model.
    For more information, see Sections \ref{subsec:IC} and \ref{subsec:Initial Evolution of the simulations}.} 
    \vspace{1mm}
    \label{fig:snapshots}
\end{figure*}

\vspace{-1mm}
\subsection{Simulation Suite and Initial Relaxation}
\label{subsec:Initial Evolution of the simulations}

We have performed a suite of simulations to investigate the galaxy--NSC--MBH co-evolution and the effects of our TDA model on the growth of the MBH (as listed in Table  \ref{table: runs}).
As discussed in Section \ref{subsec:How to compute TDR}, the two prominent parameters that the user needs to select for the sub-resolution prescription of the TDA channel are: {\it (i)} the power-law index $\gamma$ of the NSC's density profile, and {\it (ii)} the size of the NSC's stellar core, $r_\mathrm{c}$.  
The group of runs with varying $\gamma$ (from the {\tt gamma1.5} run to the {\tt gamma2.0} run) and the {\tt NoTDA1} run is called ``{\it Set A}''. 
Another group of runs with varying $r_\mathrm{c}$ (from the {\tt rc0.03} run to the {\tt rc0.3} run; for discussions on our $r_\mathrm{c}$ choices, see Appendix \ref{subsec: constraining TDA parameters}) and the {\tt NoTDA2} run is called ``{\it Set B}''.

\begin{figure}
    \vspace{0mm}
    \centering
    \includegraphics[width=0.98\columnwidth]{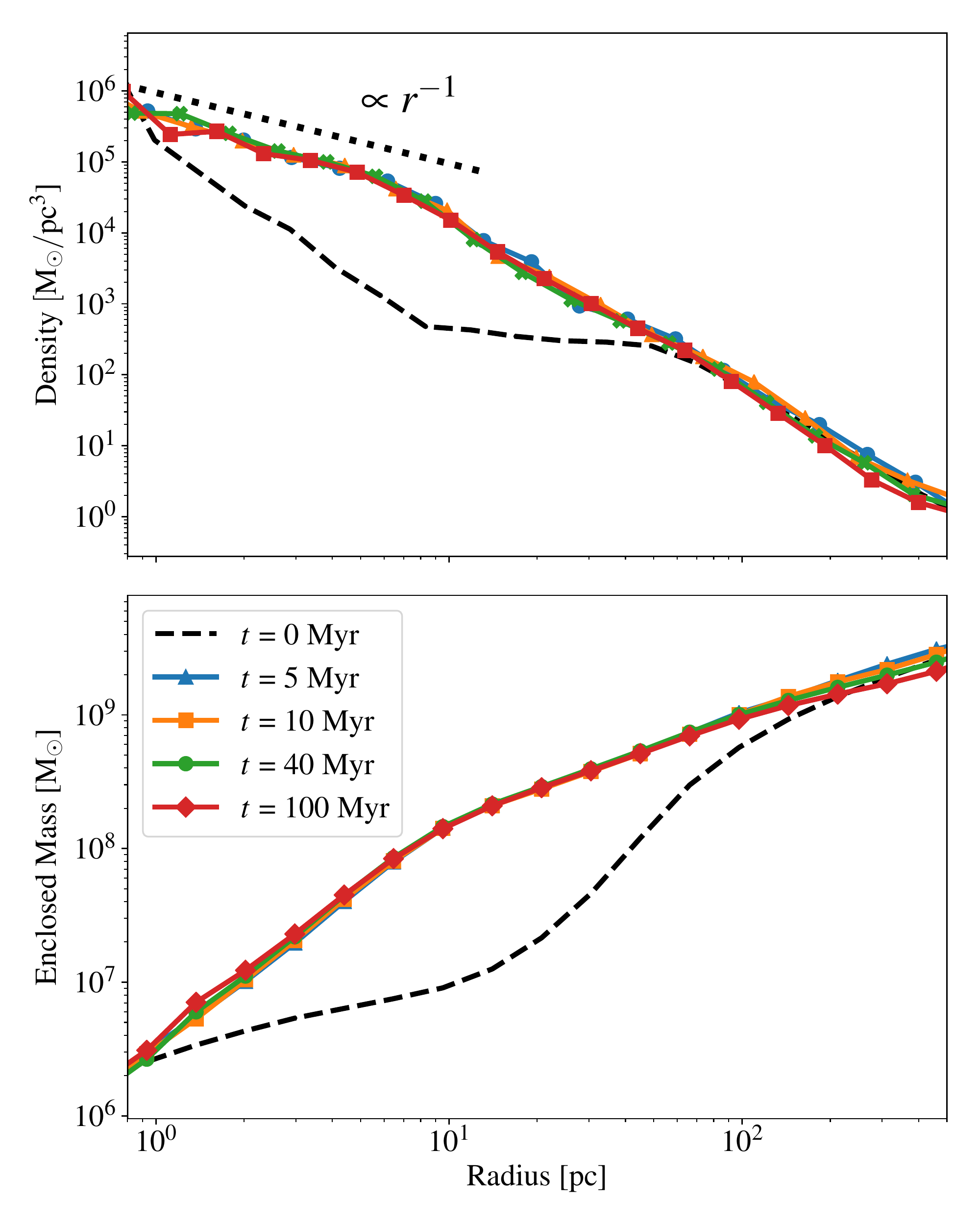}    
    \vspace{-3mm}
    \caption{The radial profiles of the stellar density ({\it top panel}) and the enclosed mass ({\it bottom panel}) from the NSC's gravitational center in the {\tt NoTDA2} run at $t = 0$, 5, 10, 40, and 100\,Myr. 
    This initial relaxation by $t \sim 5$\,Myr is due to our idealized setup following an analytic fit.
    After the initial relaxation, the profiles do not change by a significant amount. 
    For more information, see Section \ref{subsec:Initial Evolution of the simulations}.}
    \label{fig:1D Stellar density}   
\end{figure}

We first discuss the initial relaxation of the simulation in the first few Myrs using the fiducial {\tt NoTDA2} run (see Table  \ref{table: runs}). 
The gas density and the stellar distribution in the {\tt NoTDA2} run at $t = 50\,{\rm Myr}$ are shown in Figure \ref{fig:snapshots}.
Due to the relaxation of the initial density distribution, the stellar density at the galactic center changes in the first $\sim 5\,{\rm Myr}$.
Figure \ref{fig:1D Stellar density} shows the stellar density ({\it top}) and the enclosed mass profile ({\it bottom}) centered on the NSC in the {\tt NoTDA2} run at several epochs.
An increase in the stellar density between $t =0$ and 5 Myr is noticeable, while the inner density slope changes from $\gamma_\mathrm{\,\,NSC}=2.0\,$ to $\,\sim1.0$.
However, the profiles do not change significantly after 5 Myr.
The changes in the first few Myrs are not because of any astrophysical origin, but because our idealized NSC and stellar bulge in the initial condition follow the artificial analytic fits that are prone to further collapse.
Considering that there is no easy way to initialize a realistic, relaxed galaxy with all its constituents in Table \ref{table: Initial Conditions}, we may regard the $t = 5$ Myr galaxy as our {\it de facto} initial condition.

After the initial relaxation, the central stellar density inside the NSC ($\lesssim 3$\,pc from the MBH) is $\sim$$10^5\,\mathrm{M}_\odot \mathrm{pc}^{-3}$, while the 1D velocity dispersion is $\sim$$150\,\km\s^{-1}$. 
These values are consistent with what was found in massive NSCs \citep[see Figure 2 of][]{Stone2017MNRAS.467.4180S}.  
The average stellar density and velocity dispersion measured inside a sphere of $R_\mathrm{s}=0.5\,\pc \simeq \Delta x_{\rm min}$ from the MBH become $\rho_0$ and $\sigma_0$, respectively, that are used to estimate the TDR (see Section \ref{subsec:How to compute TDR}). 

\begin{figure}
    \vspace{-2mm}  
    \centering
    \includegraphics[width=0.94\columnwidth]{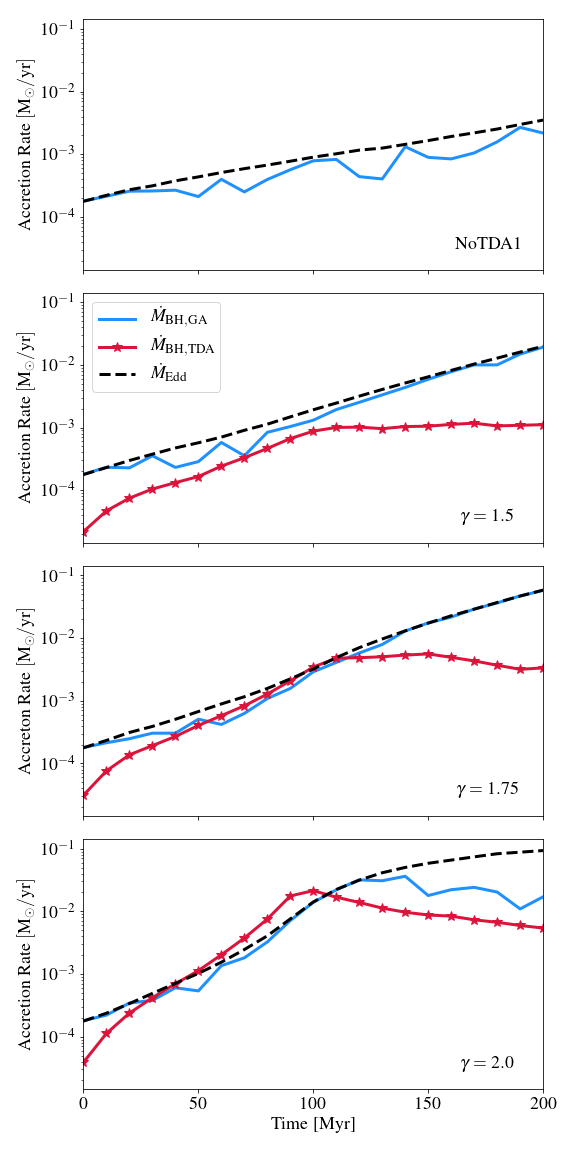}
    \vspace{-3mm}     
    \caption{The gas accretion rate (GAR $=\dot{M}_\mathrm{BH,\,GA}$; {\it blue lines}) and the tidal disruption accretion rate (TDAR $=\dot{M}_\mathrm{BH,\,TDA}$; {\it red lines}) onto the MBH in the ``{\it Set A}'' simulations (see Table  \ref{table: runs}). 
    TDAR is found to be comparable to GAR in the first $\lesssim 100$ Myr of evolution in all of the runs.  
    In the early stage of the MBH evolution, the TDAR grows with $M_\mathrm{BH}$, but it begins to saturate or even decline after $\sim$100 Myr.
    While the GAR is limited at all times by the Eddington rate (marked by {\it black dashed lines}) in all simulations, the TDAR may occasionally exceed the Eddington rate ({\it bottom panel} with $\gamma = 2.0$).
    For more information, see Section \ref{subsec:results-mbh-1}.}
    \vspace{1mm}     
    \label{fig:acc_gamma}
\end{figure}

\begin{figure}
    \vspace{-2mm}   
    \centering
    \includegraphics[width=0.94\columnwidth]{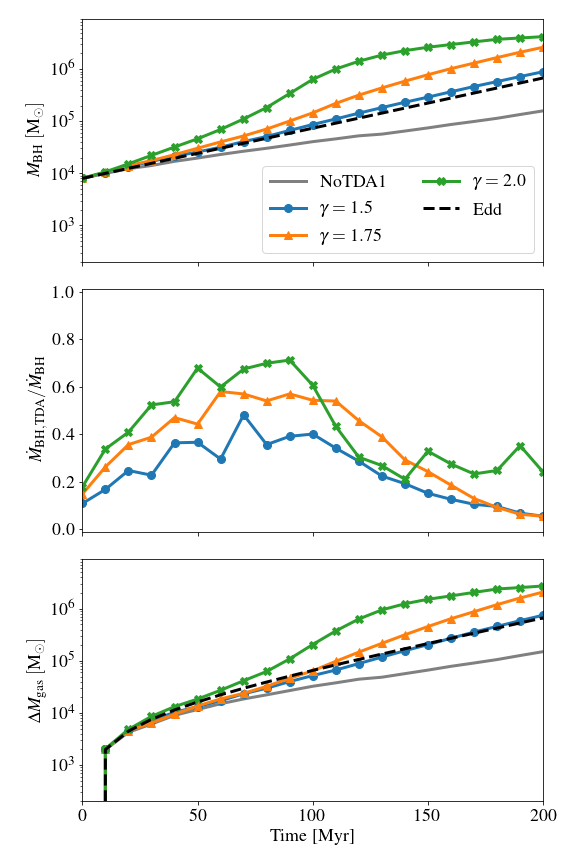}  
    \vspace{-3mm}           
    \caption{The MBH growth histories in the ``{\it Set A}'' simulations (see Table  \ref{table: runs}). 
    The {\it black dashed line}  indicates a model MBH growing at the Eddington rate. 
    {\it Top}: the MBH without TDA (the {\tt NoTDA1} run) grows only to $\sim2\times \,10^5\msun$, whereas the MBHs with TDA occasionally grow faster than the Eddington rate. 
    {\it Middle}: the ratio of TDAR to the total BH accretion rate, $\dot{M}_\mathrm{BH,\,TDA}/\dot{M}_\mathrm{BH}$. 
    The relative contribution of TDA towards the MBH's growth peaks at $ \sim$50\% around $t=50-100$ Myr. 
    {\it Bottom}: the cumulative gas mass consumed by the MBH. 
    For more information, see Section \ref{subsec:results-mbh-1}.}
    \vspace{-1mm}     
    \label{fig:Compare Gamma}
\end{figure}

\subsection{TDA's Impact on the MBH's Growth: Dependence on the Power-law Index ($\gamma$) of the NSC}
\label{subsec:results-mbh-1}

We now investigate the impact of our TDA model on the evolution of the MBH. 
In particular, by controlling the two key input parameters $\gamma$ and $r_\mathrm{c}$, we monitor how the MBH grows during the 200 Myr of evolution. 

\begin{figure}
    \vspace{-2mm}   
    \centering
    \includegraphics[width=0.94\columnwidth]{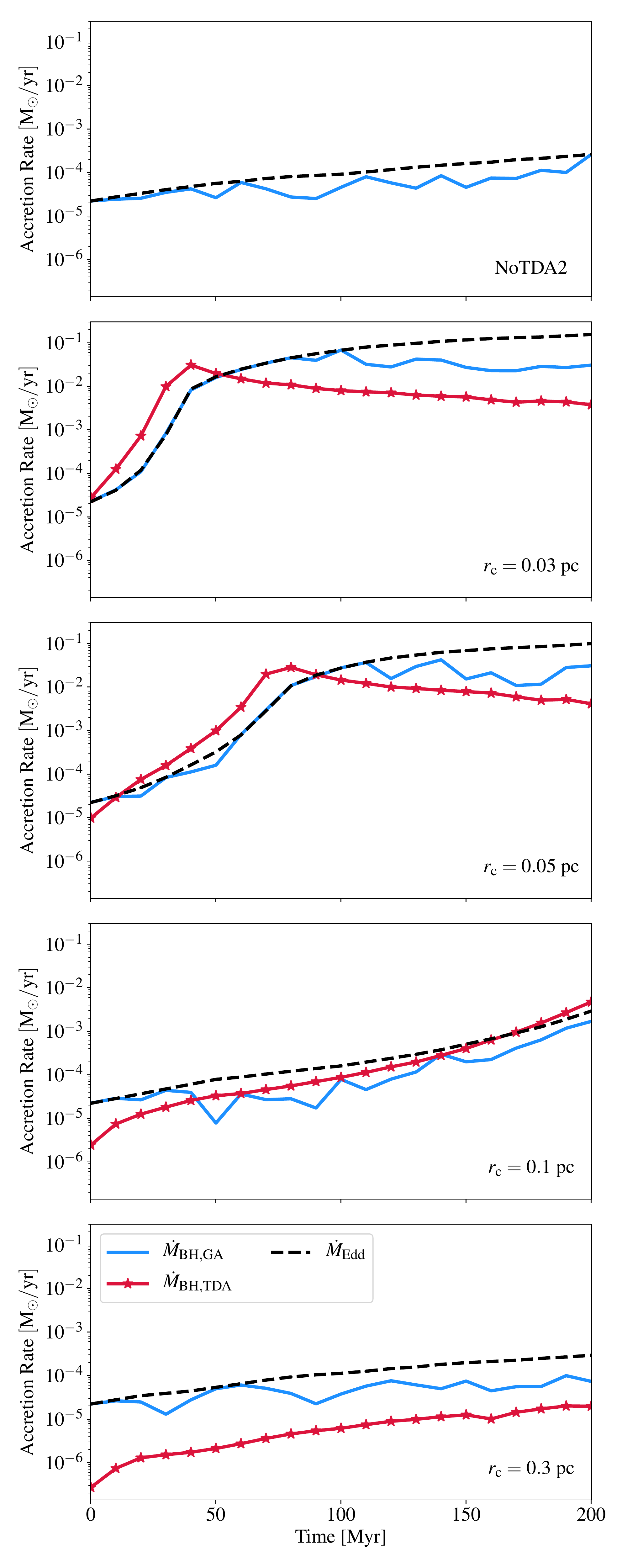}   
    \vspace{-3.5mm}
    \caption{Same as Figure \ref{fig:acc_gamma} but for the ``{\it Set B}'' simulations (see Table  \ref{table: runs}). 
    The TDAR reaches its peak earlier when a smaller $r_\mathrm{c}$ is assumed.
    For more information, see Section \ref{subsec:results-mbh-2}.}
    \vspace{-2mm}     
    \label{fig:accretion rate core size}   
\end{figure}

\begin{figure}
    \vspace{-3mm}   
    \centering
    \includegraphics[width=0.94\columnwidth]{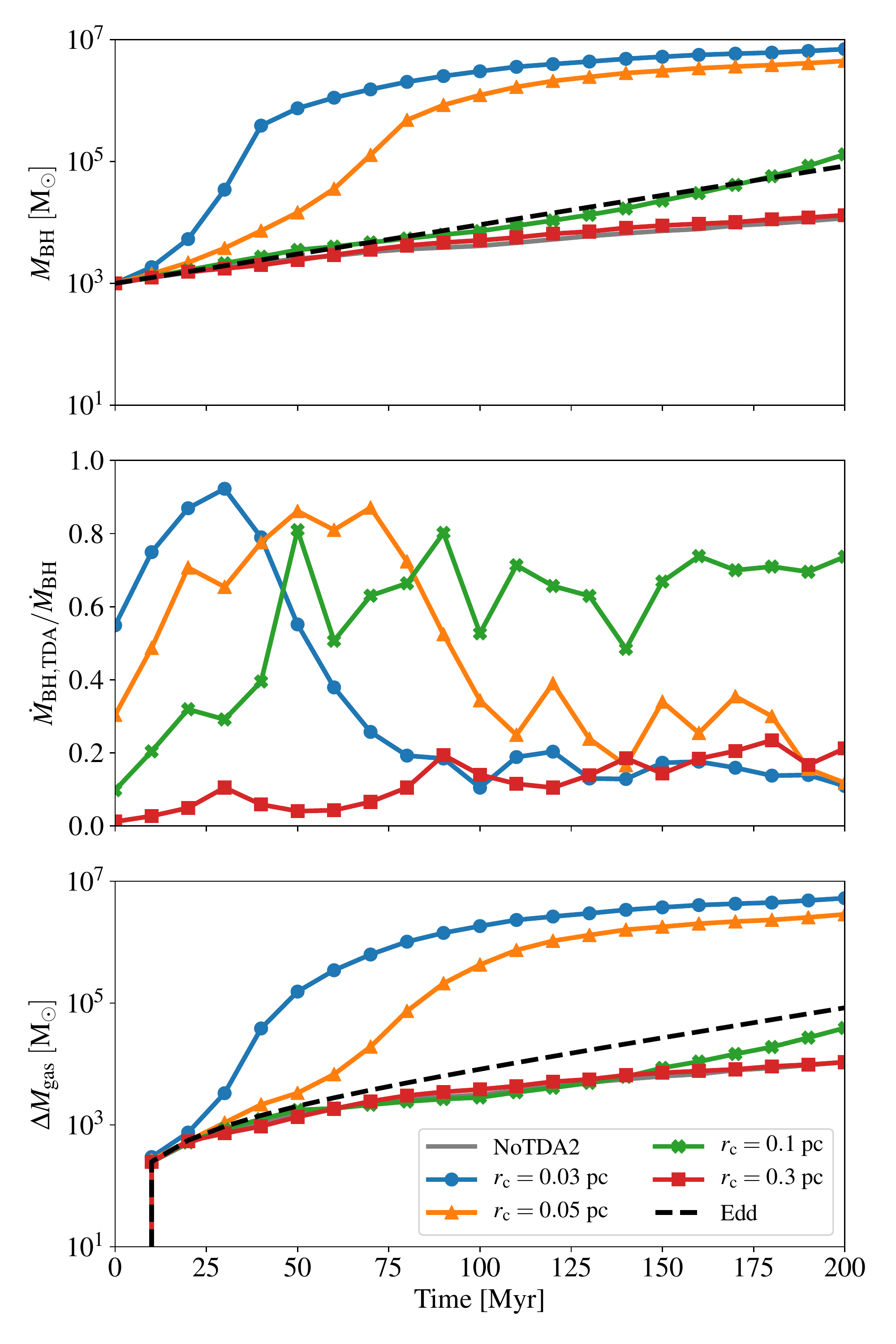}   
    \vspace{-3.5mm}
    \caption{Same as Figure \ref{fig:Compare Gamma} but for the ``{\it Set B}'' simulations (see Table  \ref{table: runs}). 
    The TDAR may (briefly) dominate over the GAR when a smaller $r_\mathrm{c}$ is assumed.
    For more information, see Section \ref{subsec:results-mbh-2}.}
    \vspace{-1mm}    
    \label{fig:Compare Core Size}
\end{figure}

First, the runs in the ``{\it Set A}'' suite (see Table  \ref{table: runs}) explore how the TDAR is affected by $\gamma$, with a MBH seed of mass $M_\mathrm{BH,\,init} = 8\times 10^3\, \mathrm{M}_\odot$. 
We are specifically interested in how the TDAR compares with the GAR.
Figure \ref{fig:acc_gamma} illustrates the GARs ({\it blue}) and TDARs ({\it red}) measured in $t=0-200$\,Myr for the ``{\it Set A}'' simulations. 
The black dashed line refers to the Eddington rate, $\dot{M}_\mathrm{BH,\,Edd}$ in Eq.(\ref{eq: mdot_gas}), corresponding to the mass of the MBH at that moment. 
What is the most worth noting is that the TDAR is comparable to GAR in the first $\lesssim 100$ Myr of evolution in all of the runs we have tested.  
In the early stage of the MBH evolution, the TDAR grows with $M_\mathrm{BH}$, but it begins to saturate or even decline after $\sim$100 Myr, which is a sign that our TDAR prescription is working as intended (see Figure \ref{fig:Expectations}).  
The TDAR becomes higher with a larger $\gamma$, as is predicted from the {\it middle} panel of Figure \ref{fig:Expectations}. 
Whereas the GAR is limited at all times by the Eddington rate, the TDAR may exceed the Eddington rate, e.g., from 60 to 100 Myr when with $\gamma = 2.0$ ({\it bottom} panel of Figure \ref{fig:acc_gamma}).  
This implies that TDA can dominate the mass supply to the MBH at a certain epoch ($M_{\rm BH} \lesssim 10^5 \, \mathrm{M}_\odot$).

In Figure \ref{fig:Compare Gamma} we further compare the MBH growth histories  in the ``{\it Set A}'' simulation suite.
By $200\,\mathrm{Myr}$, the MBH growing only via GA  reaches $\sim 2\times 10^5\msun$ in mass (the {\tt NoTDA1} run; {\it top} panel). 
In contrast, the MBHs in all of the tested runs with TDA grow faster than the Eddington rate (denoted by the black dashed line). 
These MBHs reach $\gtrsim 10^6\msun$ by 200 Myr, about an order of magnitude larger in mass than in the {\tt NoTDA1} run.  
It is worth noting that even the MBH in the {\tt gamma1.5} run --- which predicts the least TDAR among ``{\it Set A}'' (see the {\it middle} panel of Figure \ref{fig:Expectations}) --- still grows at a rate slightly higher than the Eddington rate. 
In the {\it middle} panel of Figure \ref{fig:Compare Gamma}, the ratio of TDAR to the total BH accretion rate, $\dot{M}_\mathrm{BH,\,TDA}/\dot{M}_\mathrm{BH}$, is shown.  
We see that the relative contribution of TDA towards the MBH's accretion reaches $ \sim$50\% during $50-100$ Myr ($M_{\rm BH} \lesssim 10^5 \, \mathrm{M}_\odot$).  
Indeed, for $\gamma = 1.75$ and $2.0$, the TDARs exceed the GARs in this period.
However, the ratio $\dot{M}_\mathrm{BH,\,TDA}/\dot{M}_\mathrm{BH}$ tends to decline after $100\,\mathrm{Myr}$. 
In the {\tt gamma1.5} and {\tt gamma1.75} runs, the ratio decreases to  $\lesssim 0.1$. 
In {\tt gamma2.0}, the ratio stays above 0.25 after 150 Myr, but it is mainly attributed to the stalled GA (see the {\it bottom} panel in Figure \ref{fig:acc_gamma} and Section \ref{subsec:results-sf}). 
Lastly, the {\it bottom} panel of Figure \ref{fig:Compare Gamma} depicts the cumulative gas consumption by the MBH. 
Although the GA and TDA models operate independently, TDA may still  affect GAR indirectly --- that is, when the MBH growth is boosted by TDA, the GAR in Eq.(\ref{eq: mdot_gas}) is also enhanced because of the larger MBH mass.

\subsection{TDA's Impact on the MBH's Growth: Dependence on the Core Size ($r_\mathrm{c}$) of the NSC}
\label{subsec:results-mbh-2}

We move to the results of the ``{\it Set B}'' simulations (see Table  \ref{table: runs}; with a MBH seed of $M_\mathrm{BH,\,init} = 10^3\, \mathrm{M}_\odot$) to explore how the TDAR is affected by the assumed NSC core size $r_\mathrm{c}$. 
In Figure \ref{fig:accretion rate core size}, one can see that the TDAR correlates inversely with $r_\mathrm{c}$, as our TDR prescription predicts in the {\it right} panel of Figure \ref{fig:Expectations}.  
We also find that the peak of TDAR appears earlier when a smaller $r_\mathrm{c}$ is assumed, again as expected from Figure \ref{fig:Expectations}. 
These behaviors are because a smaller $r_\mathrm{c}$ increases the TDR (see Eq.(\ref{eq: rho_c})) and at the same time reduces the transition mass $M_\mathrm{c}$  (see Eq.(\ref{eq: TDE 1}) and Section \ref{subsec:How to compute TDR}).

Figure \ref{fig:Compare Core Size} is the same as Figure \ref{fig:Compare Gamma} but for the ``{\it Set B}'' suite. 
As in Figure \ref{fig:Compare Gamma}, the MBH growth is significantly enhanced by TDA, especially in its early evolutionary stage. 
The relative contribution of TDA towards the MBH's accretion reaches $ \gtrsim$ 50\% when $M_{\rm BH} \lesssim 10^5 \, \mathrm{M}_\odot$.
The TDAR is greatly affected by the assumed value of $r_\mathrm{c}$. 
In the {\tt rc0.03} run, the MBH grows to $\sim 10^7\msun$, nearly 3 orders of magnitude greater than in our control run, {\tt NoTDA2}. 
By contrast, TDA makes little difference in the {\tt rc0.3} run, and the MBHs in the {\tt rc0.3} run and the {\tt NoTDA2} run reaches similar masses at $t=200$ Myr.   
This observation emphasizes the importance of the assumed characteristics of the NSC in our TDA prescription, such as $r_\mathrm{c}$ (for discussion on our $r_\mathrm{c}$ choices, see Appendix \ref{subsec: constraining TDA parameters}). 

To summarize our findings, we discover that TDA can significantly boost the growth of the seed MBH in most of the runs we tested with reasonable parameter choices for the NSC's structure, $\gamma$ and $r_\mathrm{c}$. 
It is especially true in the early phase of the MBH's growth up to $M_{\rm BH} \lesssim 10^5 \, \mathrm{M}_\odot$.   
During this phase, the $\dot{M}_\mathrm{BH,\,TDA}$ grows with $M_\mathrm{BH}$. 
After $\dot{M}_\mathrm{BH,\,TDA}$ reaches its peak, however, it begins to decline as $M_\mathrm{BH}$ grows, so does the contribution of TDA towards the MBH's growth (for all the run with TDA in Table \ref{table: runs} except {\tt rc0.1} and {\tt rc0.3}). 
In this later phase, GA becomes the main mass supplier to the MBH, in agreement with the conventional MBH growth model in a galaxy-scale numerical experiment.

\subsection{TDA's Impact on the MBH Host Galaxy: Star Formation, Morphology}
\label{subsec:results-sf}

The TDA model we have tested describes an interaction between the MBH particle and its neighboring star particles.  
Yet, TDA's influence reaches beyond this simplistic interaction.  
In Section \ref{subsec:results-mbh-2}, we discussed that TDA may indirectly affect the GAR by boosting the MBH's mass.  
In this section, we consider TDA's impact on other aspects of the MBH host galaxy, such as its star formation history and morphology.  

\begin{figure}
    \vspace{-2mm}    
    \centering
    \includegraphics[width=\columnwidth]{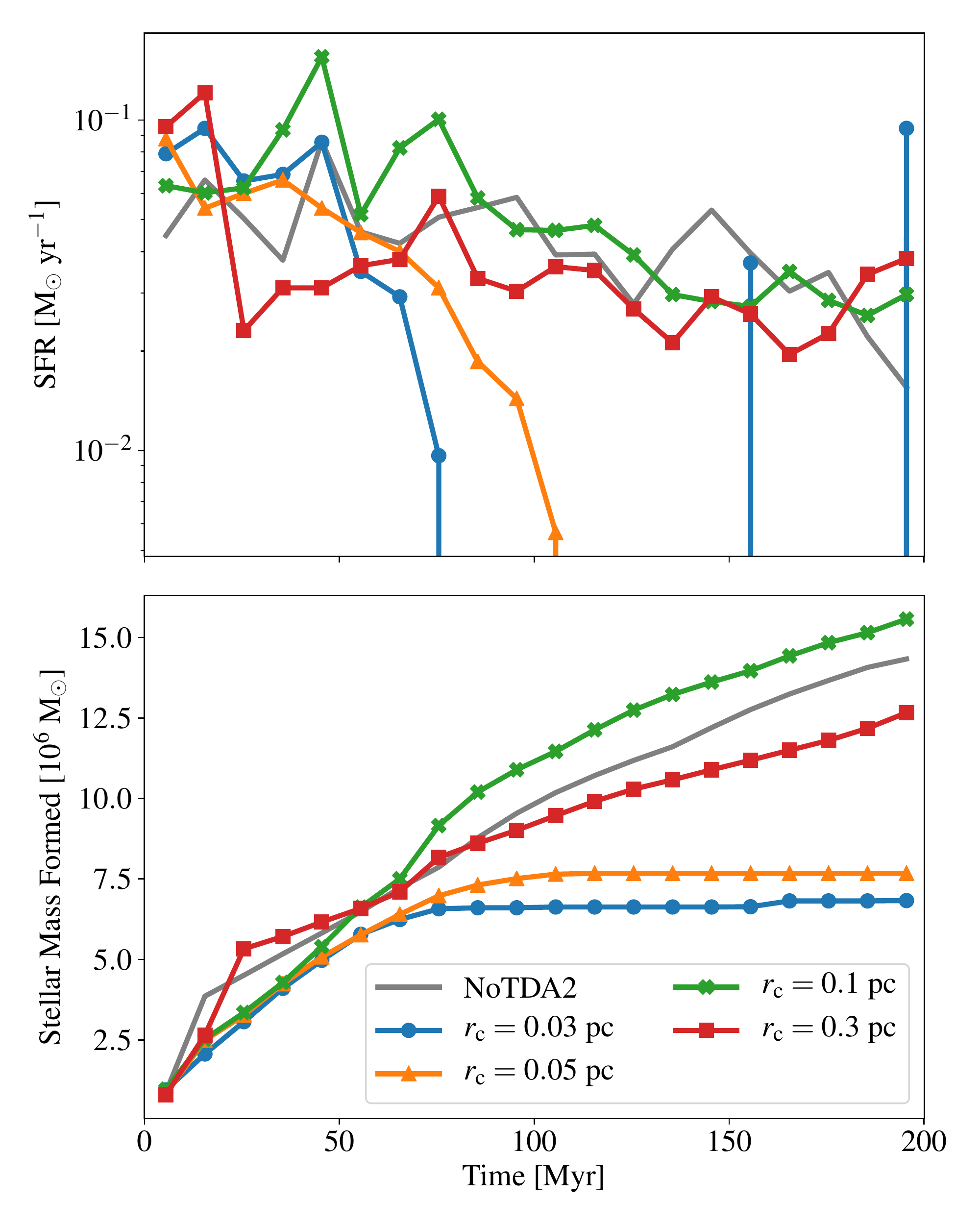}    
    \vspace{-6mm}    
    \caption{The star formation histories (SFHs) of the ``{\it Set B}'' simulations (see Table  \ref{table: runs}).  
    {\it Top}: star formation rates (SFRs). 
    {\it Bottom}: Total masses of newly-formed stars in the simulation.  
    In the {\tt rc0.03} and {\tt rc0.05} run, star formation halts after $t\sim 100\,\mathrm{Myr}$. 
    In these two runs, the total masses of the newly-formed stars by $t \sim 200\,\mathrm{Myr}$ are only half of those in the other runs.}
    \vspace{-1mm}
    \label{fig:SFR Compare}
\end{figure}

We first examine the star formation history (SFH) to study this effect. 
Figure \ref{fig:SFR Compare} shows the star formation rates (SFRs) and the cumulative masses of newly-formed star particles as functions of time for the ``{\it Set B}'' suite. 
In all the runs, stars form at a rate of $\gtrsim 0.03\,\mathrm{M}_\odot \mathrm{yr}^{-1}$ before $t \sim 50\,\mathrm{Myr}$. 
However, in the {\tt rc0.03} and {\tt rc0.05} run in which TDA is more active than others, star formation is quenched after $\sim$100 Myr.   
As a result, the cumulative masses of the newly-formed stars by $t \sim 200\,\mathrm{Myr}$ in the two runs are $\sim 7\times 10^6\msun$, while that of the other runs is $\gtrsim 1.3\times 10^6\msun$  ({\tt rc0.1}, {\tt rc0.3}, {\tt NoTDA2}). 
This mass gap is approximately the same as the difference in the cumulative gas mass consumed by the MBH during the same period (see the {\it bottom} panel of Figure \ref{fig:Compare Core Size}). 
It implies that the suppressed star formation in  the {\tt rc0.03} and {\tt rc0.05} run is likely due to the increased gas consumption by the MBH with efficient TDA. 

\begin{figure}
    \hspace*{-7mm} 
    \centering
    \includegraphics[width=1.15\columnwidth]{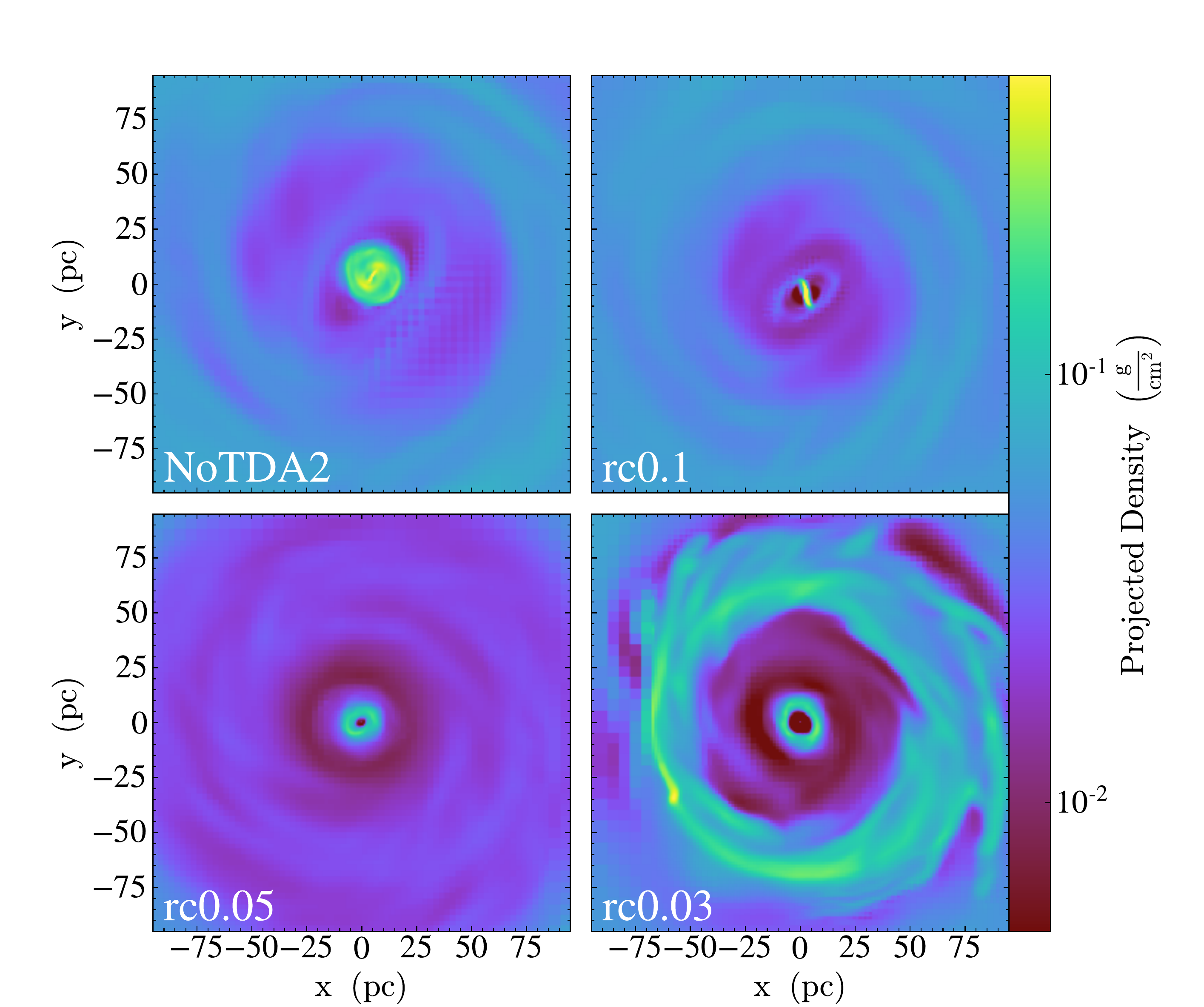}    
    \vspace{-4mm}
    \caption{The projected gas densities of the ``{\it Set B}'' simulations (see Table  \ref{table: runs}) centered on the location of the MBH at T = 150\,Myr. 
    In the {\tt NoTDA2} run, the concentrated gas around the MBH is noticeable. 
    In contrast, once we include TDA, the gas density around the black hole decreases. 
    In the {\tt rc0.03} run, for example, one can see a clear cavity (of size $\sim$ a few pc) in the vicinity of the MBH.}
    \vspace{-1mm}
    \label{fig:gas projection}
\end{figure}

When TDA is most active, the intensified gas consumption by the MBH may also reduce the gas density near the MBH.  
Sometimes it creates a visible cavity in gas in the vicinity of the MBH. 
Figure \ref{fig:gas projection} shows the projected gas densities of the ``{\it Set B}'' simulations at $t$ = 150\,Myr.
When compared with the {\tt NoTDA2} run, cavities in gas density around the MBH (of sizes $\sim$ a few pc) are pronounced in all other runs with TDA.  
One may also see a trend that the cavity size increases with decreasing $r_\mathrm{c}$. 
This morphological change is because the massive MBH --- growing expeditiously with both GA and TDA --- could consume its neighboring gas faster than being replenished by gas inflows.  
One could argue that this gas cavity has stalled the GAR in some epochs; e.g., see the {\it bottom} panel of Figure \ref{fig:acc_gamma} for the {\tt gamma2.0} run, and the {\it 2nd}/{\it 3rd} panel of Figure \ref{fig:accretion rate core size} for the {\tt rc0.03}/{\tt rc0.05} run.  
In these runs, the GAR is significantly lower than the Eddington rate after $\sim 150\,\mathrm{Myr}$ due to the lack of gas supply around the MBH.

\vspace{2mm}

\section{Discussion}
\label{sec:Discussion}

\subsection{Comparison With Observations and Future Observational Prospects}
\label{subsec: Observational Prospect}

In our simulations, we find that the TDR often reaches $10^{-2}-10^{-3}\,\mathrm{yr}^{-1}$.
In contrast, the estimates based on the observed TDE samples give $\sim 10^{-5}\yr^{-1}\,\mathrm{galaxy}^{-1}$ \citep[e.g.,][]{Donley2002AJ....124.1308D, Esquej2008A&A...489..543E, Gezari2009ApJ...698.1367G, van2014ApJ...792...53V, Holoien2016MNRAS.463.3813H}.
We however note that the actual TDR would likely be higher than $\sim 10^{-5}\yr^{-1}\,\mathrm{galaxy}^{-1}$, given that the current survey selection technique may be missing some population of TDEs and that there may be an error in the assumed BH mass function in the lower-mass galaxies \citep{Gezari2021ARA&A..59...21G}.
It should also be noted that the theorists have predicted the rates $\sim 10^{-4}-10^{-3}\yr^{-1}\,\mathrm{galaxy}^{-1}$ in the local universe, a value higher than the aforementioned observed values \citep[e.g.,][see also \citealt{van2018ASPC..517..737V}]{Magorrian1999MNRAS.309..447M, Wang2004ApJ...600..149W}.
%\cite{van2018ASPC..517..737V} showed that the observation could reproduce the expectation by the theory.
%More importantly, the expected rate above is the {\it average} value in the {\it local} galaxies.
Furthermore, the TDR can be greater at high redshift than in the local universe \citep{Kochanek2016MNRAS.461..371K}.

The growth of a MBH by TDA may be verified by the observations in the near future.
Recently, \cite{Baldassare2022ApJ...929...84B} presented an observational test of \cite{Stone2017MNRAS.467.4180S}'s work.
They found that the NSCs with velocity dispersions above $40\,\,\mathrm{km}\,\mathrm{s}^{-1}$ (a threshold suggested by \citealt{Miller2012ApJ...755...81M}) are twice as likely to contain a MBH.
In addition, measurements of the spin distribution of MBHs may determine the contribution of TDA on the growth of a MBH \citep{Zhang2019ApJ...877..143Z}.
TDE flares at $z \gtrsim 3$ could also be observed \citep{Padmanabhan2021A&A...656A..47P}.

\subsection{Limitations of Our Model and Future Work}
\label{subsec: limitations}

Our model includes several simplifying assumptions to estimate the TDAR.
Here we discuss how lifting these assumption would change our findings, and present future work.

\begin{enumerate}
    \item {\it Mass function of stars in the NSC}: 
    We have assumed the masses of stars in the NSC are identically $m_\star=0.7 \msun$ (see Sections \ref{subsec:power-law} and \ref{subsec:TDA implementation} and Eq.(\ref{eq: deltaM_BH})).  
    In reality, a NSC consists of stars with a wide range of masses. 
    In such a stellar system, the equipartition principle entails that more massive stars migrate inward.
    This implies that the average stellar mass at the center of the NSC can be higher than $m_\star=0.7 \msun$.
    Then, the relaxation time, $t_\mathrm{\,R}$ in Eq.(\ref{eq: TDE WM}), at the center of the NSC would become shorter than what our model expects. 
    Correspondingly, the TDR would be higher.\footnote{The change in $m_\star$ may also entail a steeper power-law index, and thus a change in the TDAR \citep{Bahcall1977ApJ...216..883B}.}
    
    \item {\it Fraction of the tidal debris that accretes to the MBH}: 
    We have assumed that after TDEs the fraction $f$ of stellar debris that eventually accretes to the MBH is 1.0 (see Section \ref{subsec:TDA implementation} and Eq.(\ref{eq: deltaM_BH})). 
    However, it is likely that not all the debris falls onto the MBH. 
    In typical TDEs, approximately a half of the disrupted mass is bound to the compact object \citep{Rees1988Natur.333..523R, Evans1989ApJ...346L..13E}. 
    More materials may become unbound if an energetic shock is produced by the infalling gas, making the actual accretion efficiency $f$ lower than unity \citep[e.g.,][]{Strubbe2009MNRAS.400.2070S}.\footnote{Another consideration may lower the fraction $f$.  Studies on super-critical accretion show that after a TDE, the bound material eventually returns to the pericenter at a super-Eddington rate \citep[e.g.,][]{Evans1989ApJ...346L..13E, Strubbe2009MNRAS.400.2070S} and then feeds the BH \citep[e.g.][]{Ohsuga2005ApJ...628..368O}.  When super-Eddington accretion is allowed, the typical timescale for this process is $\sim$days, and most of the bound material can be consumed by the BH before the next TDE.  However, if the accretion is limited by the Eddington rate, the next TDE will occur {\it before} all the bound debris returns to the BH.  In this case, the actual accretion efficiency could be even lower.}
    
    \item {\it MBH binaries}: 
    Our TDR estimates are based on the assumption that the NSC follows a power-law density profile with an index $\gamma$ (see Sections \ref{subsec:power-law}). 
    In reality, the assumption may not be valid.  
    For example, a second MBH spiraling into the NSC's center may perturb the stellar density significantly \citep[e.g.,][]{Merritt2006ApJ...648..890M}.
    In such a case, a different model would be needed to compute TDRs of MBH binaries \citep[e.g.,][]{Li2017ApJ...834..195L, Li2019ApJ...883..132L}. 
    During the close encounter of two MBHs the $\dot{M}_{\mathrm{BH,\,TDA}}$ may  significantly  increase, albeit temporarily, for two reasons. 
    First, strong perturbation from the MBH companion enhances the loss cone feeding. 
    Second, more stars populate the loss cone via 3-body interaction in the triaxial stellar distribution during this phase.
    These processes cannot be properly captured by our present model.
    Therefore, more sophisticated TDA model will be required in the future.

    \item {\it MBH-BH mergers}:
    Our work focuses on the MBH growth by tidal disruption of stars. 
    But, stellar mass BHs may also contribute to the (birth and) growth of the MBH.
    \citet{Antonini2019MNRAS.486.5008A} pointed out that BHs with masses $\sim 100\,\msun$ can grow in a star cluster with high density ($\gtrsim 10^5\msun\pc^{-3}$) and high escape speed ($\gtrsim 300\km\s^{-1}$). 
    \citet{Hong2018MNRAS.480.5645H, Hong2020MNRAS.498.4287H} showed that IMBHs with masses $\gtrsim 10^4\msun$ can grow by repeated mergers of stellar BHs within $12\,\mathrm{Gyr}$. 
    However, this channel is likely to be suppressed as MBH grows, as stellar mass BHs are removed by mergers, or  as they run away from the galactic center \citep{Miller2012ApJ...755...81M, Hong2020MNRAS.498.4287H}. 
        
    \item {\it Tidal captures of stars}: 
    A tidal impulse during a close encounter between a compact object and a star may help to form a binary system \citep{Fabian1975MNRAS.172P..15F}. 
    This process is called a tidal capture (TC). 
    Studies have suggested that a stellar mass black hole in a dense stellar cluster may grow to an IMBH by successively capturing nearby stars \citep[e.g.,][]{Miller2012ApJ...755...81M, Stone2017MNRAS.467.4180S}. 
    However, the TC is likely to be deactivated for BHs with masses $\gtrsim 10^3 \msun$  \citep{Stone2017MNRAS.467.4180S}. 
    Then the TDA problem can be dealt with using the classical loss cone theory adopted in our model. 
    Since the initial masses of MBH seeds in our simulations are above $10^3\msun$, we choose not to consider the TCs this time.
    
    \item {\it Resolved dynamics of the NSC}: 
    When we determine the stellar profile near the MBH, we take a sub-resolution approach by assuming a power-law profile with an index $\gamma$ irrespective of the profile found in the simulation (see Sections \ref{subsec:power-law} and \ref{subsec:How to compute TDR}, and footnote \ref{different_gammas}).
    On the other hand, \cite{Pfister2021MNRAS.500.3944P} selects $\gamma$ by computing stellar masses within two spheres of radii $2\Delta x_\mathrm{min}$ and $4\Delta x_\mathrm{min}$.
    Yet, extrapolating the stellar density from the scale of $\sim\Delta x_\mathrm{min}$ down to $\sim r_\mathrm{c}$ may yield an inaccurate $\gamma$ value when the MBH's gravitational influence is not well resolved. 
    Thus, one may say either approach has its own limitations.  
    In our future work, we plan to resolve the NSC with e.g., a direct $N$-body routine that subcycles in a hydrodynamic calculation.  
    
    \item {\it Gaseous dynamical friction}:
    A massive perturber in a gaseous medium loses its angular momentum and sinks into the center due to dynamical friction. 
    Dynamical friction may bring massive stars far outside into the NSC center. 
    Further, \cite{Boco2020ApJ...891...94B, Boco2021JCAP...10..035B} argued that dynamical friction can drive multiple mergers, thus making MBHs to grow to $\sim 10^{4-6}\msun$ in $\sim 10$ Myr.
    Higher-resolution hydrodynamic calculations may thus provide  more accurate BH growth rates.
\end{enumerate}

The prediction of the MBH growth history based on our simulations should be interpreted with caution due to the limitations above. 
Nonetheless, our sub-resolution prescription for TDA in a high-resolution galaxy-scale hydrodynamic simulation is the first step towards understanding the possible contribution of TDA to the MBH growth.

%\vspace{10mm}
\section{Conclusion}
\label{sec:Conclusion}

We have introduced a model of MBHs' growth via TDA and quantified its effects on the MBH evolution using high-resolution AMR simulations. 
Despite its possible contribution, the role of TDA towards the MBH growth has been overlooked in most galaxy simulations. 
GA may dominate the accretion to the MBHs with masses $\gtrsim 10^6\,\mathrm{M}_\odot$, but TDA may significantly boost the growth of the MBH seeds with masses $\lesssim10^{5}\,\mathrm{M}_\odot$. 
Yet, it is nontrivial to attain sufficient resolution to describe the interactions between stars and the MBH in a galaxy-scale simulation, due to the limited computational resources (Section \ref{sec: intro}). 
To tackle the challenge, we have built a sub-resolution prescription adopting a statistical approach with a few simplifying assumptions (Section \ref{sec: Tidal Disruption Accretion Model}). 
With the new model that considers both GA and TDA in a galaxy-scale hydrodynamic simulation (Section \ref{sec:Simulations}), we have acquired three main results (Section \ref{sec:Result}).

\begin{enumerate}
    \item TDA significantly enhances the MBH's growth. 
    In some simulations, a MBH seed grows rapidly from  $10^3\msun$ to $\gtrsim 10^6\,\mathrm{M}_\odot$ in 200\,Myrs (Figures \ref{fig:acc_gamma} and \ref{fig:accretion rate core size}). 
    The growth rate is more than an order of magnitude higher than in the run where a MBH grows only via Eddington-limited GA.
    In general, the MBH grows faster if its host NSC is assumed to have a higher power-law index ($\gamma$) and a smaller core size ($r_\mathrm{c}$). 
    \item TDA mainly contributes to the early growth of MBH, from $10^{3-4}\msun$ to $\lesssim10^{5}\,\mathrm{M}_\odot$. 
    In most tested runs, the relative contribution of TDA towards the MBH's accretion reaches $ \gtrsim$ 50\% when $M_{\rm BH} \lesssim 10^5 \, \mathrm{M}_\odot$ (Figures \ref{fig:Compare Gamma} and \ref{fig:Compare Core Size}).
    As the MBH grows, TDA becomes sub-dominant while the later evolution is driven by GA. 
    \item We also find that the star formation around the MBH is suppressed when TDA is most active, because the massive MBH growing by both GA and TDA could rapidly consume its neighboring gas. 
    A cavity in gas (of size $\sim$ a few pc) is sometimes visible near the MBH.
\end{enumerate}

While our experiments are ideal to see the relative contribution of TDA versus GA in a simplified setting (e.g., the NSC with a power-law profile), the TDAR in reality may be different from our estimates.  
We also note that our sub-resolution TDA model in simulations behaves in the exact way that it is designed (as a function of $M_\mathrm{BH}, \rho_0, \sigma_0, \gamma,$ and $r_{\rm c}$), and the model is not intended to probe how the TDAR changes as a result of the detailed stellar dynamics in the NSC resolved in the simulation.  
And yet, our study is sufficient to show the possible contribution of TDA to the rapid growth of MBHs. 
Our calculations demonstrate the need to consider different channels of MBH accretion that may provide clues for the existence of supermassive black holes at high redshifts.   
We also emphasize that this is the first step towards exploring TDA for MBHs. 
In our future work, we aim  not only to advance our TDA model, but also to test it in various environments. 
Further improvements to our model will make our prediction more sophisticated and reliable.

%\section*{Acknowledgements}
\vspace{2mm}

We are deeply grateful to Hyung Mok Lee and Jong-Hak Woo for their invaluable advice during this work. 
We thank the anonymous referee for providing us with insightful comments that improved this article.
We also thank Vivienne Baldassare, Lumen Boco, Jongsuk Hong, Hamsa Padmanabhan, Vivienne Baldassare, Hugo Pfister, and Xiaoxia Zhang for their detailed comments on our manuscript, and Yongseok Jo, Seoyoung Kim, and Ahram Lee for their advice during the early phase of this work.
Ji-hoon Kim acknowledges support by Samsung Science and Technology Foundation under Project Number SSTF-BA1802-04, and by the POSCO Science Fellowship of POSCO TJ Park Foundation.  
His work was also supported by the National Institute of Supercomputing and Network/Korea Institute of Science and Technology Information with supercomputing resources including technical support, grants KSC-2020-CRE-0219 and KSC-2021-CRE-0442.
The publicly available {\sc Enzo} and {\tt yt} codes used in this work are the products of collaborative efforts by many independent scientists from numerous institutions around the world.  
Their commitment to open science has helped make this work possible. 

\appendix

\section{The Mass of A MBH Seed in Our Simulations}
\label{subsec: constraining mbh seed mass}
 
Recall that, to simplify the TDR estimates, we have assumed that the MBH stays at the center of the NSC without ``wandering''  (see Section \ref{subsec:power-law}). 
This assumption must hold true during the simulation to make our TDA model self-consistent.  
This consideration yields a requirement for the initial MBH seed mass in our runs, $M_\mathrm{BH, \,init}$ (see Section \ref{subsec:IC}).

To show this, we start by imposing a criterion
\begin{equation}
    r_\mathrm{wan} < r_\mathrm{\,infl}
\label{eq: BH wander criteria}
\end{equation} 
where $r_\mathrm{wan}$ is the {\it wandering radius} of the MBH that can be estimated from
\begin{equation}
    \frac{v_\mathrm{BH}^2}{r_\mathrm{wan}} = \frac{G M_\star\left(r_\mathrm{wan}\right)}{r_\mathrm{wan}^2},
\label{eq: BH speed}
\end{equation}
while the MBH's {\it radius of influence} \citep{Peebles1972GReGr...3...63P} is written as
\begin{equation}
    r_\mathrm{\,infl} = \frac{G M_\mathrm{BH}}{\sigma_{\rm NSC}^2}\,.
\label{eq: BH radius of influence}
\end{equation}
In both Eqs.(\ref{eq: BH speed}) and (\ref{eq: BH radius of influence}), a nearly isothermal stellar distribution is assumed (i.e., $\gamma = 2.0$ in Eq.(\ref{eq: power-law with core})).  
Therefore, the stellar mass enclosed in $r_\mathrm{wan}$ is $M_\star\left(r_\mathrm{wan}\right) = (4\pi/3) \rho_\mathrm{NSC} r_\mathrm{wan}^3$.   
Because of the isothermal sphere assumption (with $\rho_\mathrm{c} = \rho_\mathrm{NSC}$), we can then write
\begin{equation}
    r_{\rm wan}^2 = \frac{v_\mathrm{BH}^2}{(4\pi/3)G \rho_\mathrm{NSC}} \,\,\,\,\,\,\,\, {\rm and} \,\,\,\,\,\,\,\, r_{\rm c}^2 = \frac{\sigma_\mathrm{NSC}^2}{(4\pi/3)G \rho_\mathrm{NSC}}
\end{equation}
to which the equipartition principle, $M_\mathrm{BH} v_\mathrm{BH}^2 \simeq 3m_\star \sigma_\mathrm{NSC}^2$, is applied to get 
\begin{equation}
    r_\mathrm{wan} \simeq r_\mathrm{c}\left(\frac{m_\star}{M_\mathrm{BH}}\right)^{{1 \over 2}}.
\label{eq: BH wander radius}
\end{equation}
Finally, plugging Eqs.(\ref{eq: BH radius of influence}) and (\ref{eq: BH wander radius}) into Eq.(\ref{eq: BH wander criteria}), we obtain 
\begin{align}
    &M_\mathrm{BH,\,init} > \left(
    \frac{r_\mathrm{c}^2 \sigma_{\rm NSC}^4 m_\star}{G^2}
    \right)^{\frac{1}{3}} \nonumber \\
    \approx 
    &\,\,3\times10^3\msun\times\left( \frac{r_\mathrm{c}}{0.1\pc}\right)^{\frac{2}{3}}\left(\frac{\sigma_{\rm NSC}}{100\km\s^{-1}}\right)^{\frac{4}{3}}\left(\frac{m_\star}{0.7\msun}\right)^{\frac{1}{3}}
\end{align}
which gives a lower limit of $M_\mathrm{BH,\, init}$ that makes the MBH stationary and our TDA model self-consistent (see Section \ref{subsec:power-law}). 
This motivates our choices of $M_\mathrm{BH,\, init}$ in Section \ref{subsec:IC}. 

\section{The Range of the NSC's Core Size in Our Model}
\label{subsec: constraining TDA parameters}

Our TDA prescription computes $\dot{M}_\mathrm{BH,\,TDA}$ based on the user-defined parameters, such as the power-law index $\gamma$ of the NSC's density profile, and the size of its stellar core, $r_\mathrm{c}$ (see Section \ref{subsec:How to compute TDR}). 
Therefore, it is crucial to provide the model with well-constrained parameters in simulations with TDA, in order to avoid an unrealistic MBH growth scenario. 

In particular, given that a small change in $r_\mathrm{c}$ results in a sizable change in the TDAR, it is essential to constrain $r_\mathrm{c}$ to be adopted for the TDA model. 
\citet[][see their Figure 3]{Stone2017MNRAS.467.4180S} made a 2D Gaussian fit between $\overline{\sigma}_{\rm NSC}$ and $r_\mathrm{c}$ using observational data \citep[e.g.,][]{Boker2004AJ....127..105B, Cote2006ApJS..165...57C, Boker2014MNRAS.441.3570G} --- where $\overline{\sigma}_{\rm NSC} \equiv \sqrt{GM_{\rm NSC}/(3r_{\rm \,NSC})}$ is the average 1D velocity dispersion in the NSC --- and found a relation $r_\mathrm{c} \propto (\overline{\sigma}_{\rm NSC})^{-1.3}$. 
It means that  the NSC with greater $\overline{\sigma}_{\rm NSC}$ tends to have a smaller $r_\mathrm{c}$, thus a denser NSC core and more active TDA. 
From their Figure 3 one can also observe that $r_\mathrm{c}$ is approximately within $[ 0.1, \,1] \pc$ for $\overline{\sigma}_{\rm NSC} \sim 100\, \km \s^{-1}$.  
This motivated our chosen $r_\mathrm{c}$ range for simulations, $[ 0.03, \,0.3] \pc$ (see Table \ref{table: runs}).  
However, because the $\overline{\sigma}_{\rm NSC}-r_\mathrm{c}$ relation has a large scatter, it is difficult to use the  relation to constrain $r_\mathrm{c}$ exactly.    

%%%%%%%%%%%%%%%%%%%%%%%%%%%%%%%%%%%%%%%%%%%%%%%%%%



%\begin{thebibliography}{99}
%\end{thebibliography}

%%%%%%%%%%%%%%%%%%%%%%%%%%%%%%%%%%%%%%%%%%%%%%%%%%

%%%%%%%%%%%%%%%%% APPENDICES %%%%%%%%%%%%%%%%%%%%%




\bibliography{main}{}
\bibliographystyle{aasjournal}



\end{document}